\documentclass{jfm}
\usepackage{graphicx}
\usepackage{epstopdf, epsfig}
\usepackage{amsfonts,amssymb,amsmath,mathbbol}   

\usepackage{xcolor} 
\usepackage{float}
\usepackage{placeins}
\usepackage{physics}

\def\We{{\it We}}
\def\d{{\rm d}}

\shorttitle{Liquid jet breakup in a crossflow airstream}
\shortauthor{G. P. Majji, A. J. Tudu, L. D. Chandrala and K. C. Sahu}

\title{Dynamics of liquid jet breakup and droplet size distribution in a crossflow airstream}

\author{Gowri Padmaja Majji\aff{1}, Aman John Tudu\aff{2}, Lakshmana Dora Chandrala\aff{1} \and Kirti Chandra Sahu\aff{2}\corresp{\email{lchandrala@mae.iith.ac.in, ksahu@che.iith.ac.in}}}

\affiliation{\aff{1} Department of Mechanical and Aerospace Engineering, Indian Institute of Technology Hyderabad, Kandi - 502 284, Sangareddy, Telangana, India \\
\aff{2}Department of Chemical Engineering, Indian Institute of Technology Hyderabad, Kandi - 502 284, Sangareddy, Telangana, India}

\begin{document}

\maketitle

\begin{abstract}
The fragmentation of a liquid jet in a crossflowing airstream arises from a complex interplay between aerodynamic forcing and interfacial instabilities acting across a wide range of length scales. We examine the spatial evolution of droplet size distributions along the jet trajectory using a combination of high-speed shadowgraphy and digital in-line holography, allowing simultaneous characterization of jet morphology and volumetric droplet size distributions. The results demonstrate a clear progression of breakup mechanisms, from near-field bag rupture to downstream rim, ligament, and node disintegration, each associated with distinct characteristic length scales. This evolution is reflected in the droplet size distributions, which broaden and become increasingly polydisperse downstream relative to the near-field distributions. By systematically varying the liquid-jet and airstream Weber numbers, we identify distinct breakup modes, and construct a regime map that delineates the column-breakup, isolated-bag-breakup, and interacting-bag regimes. Furthermore, by incorporating the dominant instability mechanisms, including Rayleigh–Taylor and Rayleigh–Plateau instabilities associated with the observed column, bag, and multi-bag breakup modes, we develop a theoretical framework for jet atomization in crossflow. The framework relates instability wavelengths to characteristic droplet sizes, and reasonably captures the experimentally observed trends. We found that stronger aerodynamic forcing induces earlier breakup, decreases the characteristic instability length scales, and generates finer droplets dispersed over a wider spatial region. Overall, the study establishes a quantitative relationship between instability-driven fragmentation dynamics and droplet size distributions, thereby connecting interfacial instabilities with measurable spray characteristics.
\end{abstract}


\section{Introduction} \label{sec:intro}

The interaction of a liquid jet with an airstream is a canonical problem in multiphase fluid dynamics, which is relevant to several natural phenomena and engineering applications, such as fuel injection in gas turbines and internal combustion engines, spray cooling, agricultural atomization, and environmental dispersion processes \citep{mahesh2013interaction,zhou2023review,traverso2023data,zhang2024review,chakraborty2025JAS,medipati2025elliptic}. 

The pioneering work of \citet{savart1833wemoire} first demonstrated that jet breakup in a quiescent ambient medium arises from the growth of interfacial perturbations that lead to disintegration. This simplified configuration has since been extensively studied \citep{eggers2008physics,kirar2024dynamics,mu2025numerical}. The breakup process culminates in droplet formation through successive stages of stretching, necking, and pinch-off \citep{borthakur2019dynamics,borthakur2017formation,kirar2024dynamics,sun2018dynamics}. Based on these studies, jet breakup is commonly classified into Rayleigh (or dripping), wind-induced, and atomization regimes, governed by the interplay of inertial, viscous, and aerodynamic forces \citep{li2019experimental,chakraborty2022effect,reitz1978atomization,richards1994dynamic,machicoane2023regimes}. In the Rayleigh regime, breakup is driven by capillary instability, producing periodic droplets, while the dripping regime occurs at very low flow rates with droplet detachment near the nozzle. As the flow velocity increases, aerodynamic effects become dominant, leading to shear-driven breakup in the wind-induced regime; at sufficiently high airflow, the jet undergoes complete atomization into a fine spray.

The dynamics become considerably more complex when a liquid jet interacts with a crossflowing airstream, which is the focus of the present study. When injected transversely into a high-speed airstream, the jet undergoes rapid bending and deformation, followed by the growth of interfacial instabilities that ultimately lead to breakup and atomization \citep{broumand2019liquid}. The resulting flow is inherently multiscale, involving strong coupling between the liquid core, interfacial waves, gas-phase vortical structures, and dispersed droplets. Despite its geometric simplicity, the system exhibits highly nonlinear behavior due to the combined effects of liquid inertia, aerodynamic forcing, viscous stresses, and surface tension \citep{lefebvre2017atomization}. In crossflow configurations, the primary governing parameters are the aerodynamic Weber number, which quantifies the relative importance of aerodynamic stresses to surface tension, and the momentum flux ratio, which characterizes the balance between jet inertia and crossflow momentum. Early experimental studies identified key features such as jet penetration length, trajectory scaling, and distinct breakup regimes \citep{becker2002breakup,varga2003initial,lee2007primary}. At low Weber numbers, the liquid column remains relatively intact and breaks up through large-scale deformation and ligament formation, often associated with column or bag-type modes \citep{wu1997breakup,wu1998spray,wang2012bag,olyaei2020experimental}. As the Weber number increases, enhanced aerodynamic forcing amplifies interfacial instabilities, with both Rayleigh–Taylor and Kelvin–Helmholtz mechanisms contributing to surface wave growth and mass stripping from the jet periphery \citep{lin1998drop,jarrahbashi2016early}. At high Weber numbers, surface tension effects become negligible, and the jet undergoes rapid shear-driven atomization, producing fine droplets and a highly dispersed spray \citep{wu1998spray}. The momentum flux ratio primarily governs jet penetration and trajectory, while also influencing breakup by regulating the exposure of the liquid column to crossflow-induced shear. Higher values promote deeper penetration and a more coherent liquid core, whereas lower values enhance deflection and accelerate atomization \citep{becker2002breakup,lee2007primary}. Additionally, variations in the effective frontal area modify aerodynamic drag and the resulting acceleration of the jet, thereby altering instability growth and transitions between breakup modes \citep{sallam2004breakup}.

In liquid jet–airstream interactions, configurations are commonly classified based on the relative orientation of gravity and the jet injection direction, namely upward and downward jet-in-crossflow. Review articles by \cite{zhang2024review,zhou2023review} provide comprehensive overviews of breakup dynamics in both configurations. While the flow is primarily governed by the aerodynamic Weber number and the momentum flux ratio, gravity introduces an additional body force that can significantly influence jet deformation and breakup by modifying the effective acceleration of the liquid column and the growth of interfacial instabilities. Its influence becomes particularly important at moderate-to-low crossflow velocities (low Weber numbers) or low momentum flux ratios, where aerodynamic forcing is weaker and the jet residence time is longer. Gravity effects are also more pronounced for larger jet diameters or denser liquids, where gravitational and aerodynamic timescales become comparable. In the upward jet-in-crossflow configuration, where the jet is injected against gravity into a transverse airstream, buoyancy opposes aerodynamic forcing, leading to enhanced deceleration, increased jet curvature, and an earlier onset of Rayleigh–Taylor-type instabilities due to adverse acceleration at the liquid–gas interface. In contrast, in the downward configuration, where the jet is injected along gravity, gravitational and aerodynamic forces act in the same direction, increasing the effective acceleration of the liquid column, promoting deformation, and altering the breakup mechanism. Consequently, gravity not only modifies jet penetration and large-scale deformation but also influences the relative roles of Rayleigh–Taylor and Kelvin–Helmholtz instabilities, as well as their coupling with gas-phase vortical structures, such as counter-rotating vortex pairs that enhance entrainment and unsteady forcing on the jet \citep{fric1994vortical,mahesh2013interaction}. The Froude number characterizes the relative importance of inertial and gravitational forces. At high Froude numbers, inertial and aerodynamic forces dominate the local momentum balance, and the direct influence of gravity on the initiation of jet breakup and the development of the primary interfacial instabilities is comparatively small. However, gravity influences the global evolution of the liquid jet through its orientation relative to the jet, thereby affecting the jet trajectory, residence time, and downstream spray development, even in the high-Froude-number regime. Consequently, the upward and downward jet-in-crossflow configurations exhibit distinct global flow evolution and spray characteristics, despite their local breakup dynamics being governed predominantly by inertial and aerodynamic forces.

Experimental investigations of the upward jet-in-crossflow configuration have identified distinct breakup regimes, ranging from column breakup to shear-driven stripping and full atomization, with transitions primarily governed by the Weber number and momentum flux ratio \citep{wu1997breakup,wu1998spray}. Subsequent studies have clarified the dominant instability mechanisms, showing that Rayleigh–Taylor modes control large-scale deformation, while Kelvin–Helmholtz instabilities drive ligament formation and droplet detachment \citep{lin1998drop,jarrahbashi2016early}. Using high-speed imaging and shadowgraphy, \cite{prakash2018liquid} further demonstrated that inlet flow conditions strongly influence jet trajectory, breakup behavior, and droplet size, with turbulent jets exhibiting enhanced atomization and altered breakup pathways. Complementary numerical simulations have revealed the multiscale nature of turbulence–interface interactions, highlighting the cascade from coherent jet deformation to dispersed spray \citep{desjardins2008high}.

A wide range of diagnostic techniques has been employed to investigate jet breakup in a crossflowing airstream \citep{medipati2025elliptic}. These include shadowgraphy, particle/droplet image analysis (PDIA), particle image velocimetry (PIV), and phase Doppler particle analysis (PDPA), each offering complementary insights into interfacial dynamics and droplet formation. High-speed shadowgraphy and backlit imaging are among the most widely used methods for capturing jet penetration, trajectory, and primary breakup structures (see \cite{prakash2018liquid} and references therein). Particle-based techniques, such as PDIA and PDPA, provide quantitative measurements of droplet size and velocity in the downstream spray. PDPA is widely used for pointwise droplet measurements \citep{bachalo1994phase,bachalo1984phase}, whereas PDIA enables spatially resolved characterization of breakup-generated droplets in jet-in-crossflow configurations \citep{boggavarapu2021secondary}. More recently, volumetric diagnostics such as digital in-line holography have emerged as powerful tools for three-dimensional spray characterization, enabling reconstruction of droplet size, position, and velocity fields. While this technique has been extensively used for single-droplet fragmentation studies \citep{guildenbecher2017characterization, shao2020machine}, its application to complex jet breakup flows remains limited. In single-droplet fragmentation, holography has enabled detailed three-dimensional measurements of droplet size distributions \citep{guildenbecher2017characterization, shao2020machine, radhakrishna2021experimental, essaidi2021aerodynamic, li2022secondary, ade2022droplet, ade2023size, ade2024prf}. Its capability is further enhanced when combined with deep-learning-based reconstruction methods, allowing spatially resolved characterization of satellite droplets \citep{shao2020machine, chakraborty2025droplet}. Using this approach, \citet{ade2023size} showed that dual-bag and retracting-bag breakup produce bimodal size distributions, whereas single-bag breakup leads to trimodal distributions. They also demonstrated that the analytical model of \citet{jackiw2022prediction} accurately predicts droplet size distributions over a wide range of Weber numbers in crossflow configurations \citep{ade2022droplet,ade2023size,ade2024prf}. However, to the best of our knowledge, only one study in the literature, by \citet{wang2022spray}, has employed high-speed digital in-line holography to investigate an upward liquid jet in crossflow, albeit restricted to the fully developed breakup region.

As the above-mentioned review indicates, the upward jet-in-crossflow configuration has been extensively studied due to its relevance in propulsion and spray applications \citep{zhang2024review,mahesh2013interaction,prakash2018liquid}. In contrast, the downward configuration has received comparatively little attention \citep{sallam2004breakup,broumand2019liquid}. Earlier studies have established the primary breakup modes and regime transitions of liquid jets in crossflow \citep{marmottant2004spray,mazallon1999primary}. Specifically, \cite{marmottant2004spray} developed a physical framework for spray formation by demonstrating that atomization is governed by the amplification of interfacial instabilities and the subsequent fragmentation cascade, thereby linking breakup mechanisms to the resulting droplet size distribution. \cite{mazallon1999primary} performed a comprehensive experimental investigation of the primary breakup of non-turbulent round liquid jets in a gas crossflow, elucidating the roles of aerodynamic forcing and liquid inertia in governing jet breakup and droplet formation. However, most previous studies have focused primarily on breakup mechanisms, jet morphology, or global spray characteristics. Furthermore, the quantitative characterisation of droplet size distributions has received comparatively little attention, despite its fundamental importance to evaporation, mixing, and combustion processes \citep{becker2002breakup,sirignano2010fluid,jenny2012modeling,villermaux2007fragmentation}. In particular, the coupled evolution of jet morphology and spatially resolved volumetric droplet size distributions in a downward jet-in-crossflow remains largely unexplored, even though the breakup dynamics and atomization characteristics differ fundamentally from those of the upward configuration. Motivated by this gap, the present study investigates the breakup dynamics and quantifies the spatial evolution of droplet size distributions in a downward liquid jet in a crossflowing airstream. To the best of our knowledge, the present study is the first to systematically investigate the spatial evolution of volumetric droplet size distributions in a liquid jet in crossflow by combining digital inline holography (DIH) measurements with instability-based theoretical modelling.

In the present study, we examine the breakup dynamics of a continuous downward liquid jet in a crossflowing airstream and characterize the spatial evolution of the resulting droplet size distributions using high-speed shadowgraphy and digital in-line holography. A central contribution of this work is the integration of advanced image processing with machine-learning-based digital in-line holographic reconstruction to obtain volumetric droplet size distributions. We show that the droplet size distribution becomes time-invariant once the flow reaches a fully developed state. By systematically varying the liquid-jet and airstream Weber numbers, we identify distinct breakup regimes and quantify crucial features, including jet trajectory, spreading, and breakup length. We establish a regime map in the liquid-jet Weber number and airstream Weber number space that delineates column breakup, isolated-bag breakup, and interacting-bag regimes. By accounting for different instability mechanisms, such as Rayleigh–Taylor (RT) and Rayleigh–Plateau (RP), to the observed breakup modes (column, bag, and multi-bag), we establish a theoretical framework for jet atomization in crossflow. Additionally, we quantify the droplet size distributions by incorporating the characteristic length scales associated with different breakup processes, such as bag, rim, and node breakup, which dominate in different spatial regions. 

The remainder of the paper is organised as follows. Section \S\ref{sec:expt} describes the experimental setup and procedures used to investigate jet breakup dynamics in a crossflowing airstream. Section \S\ref{sec:dis} presents the results and discussion. Finally, Section \S\ref{sec:conc} summarises the main findings and provides concluding remarks.

\section{Experimental setup and procedure}\label{sec:expt}

\subsection{Experimental setup}\label{subsec:exp_setup}

\begin{figure}
\centering
\includegraphics[width=1.0\textwidth]{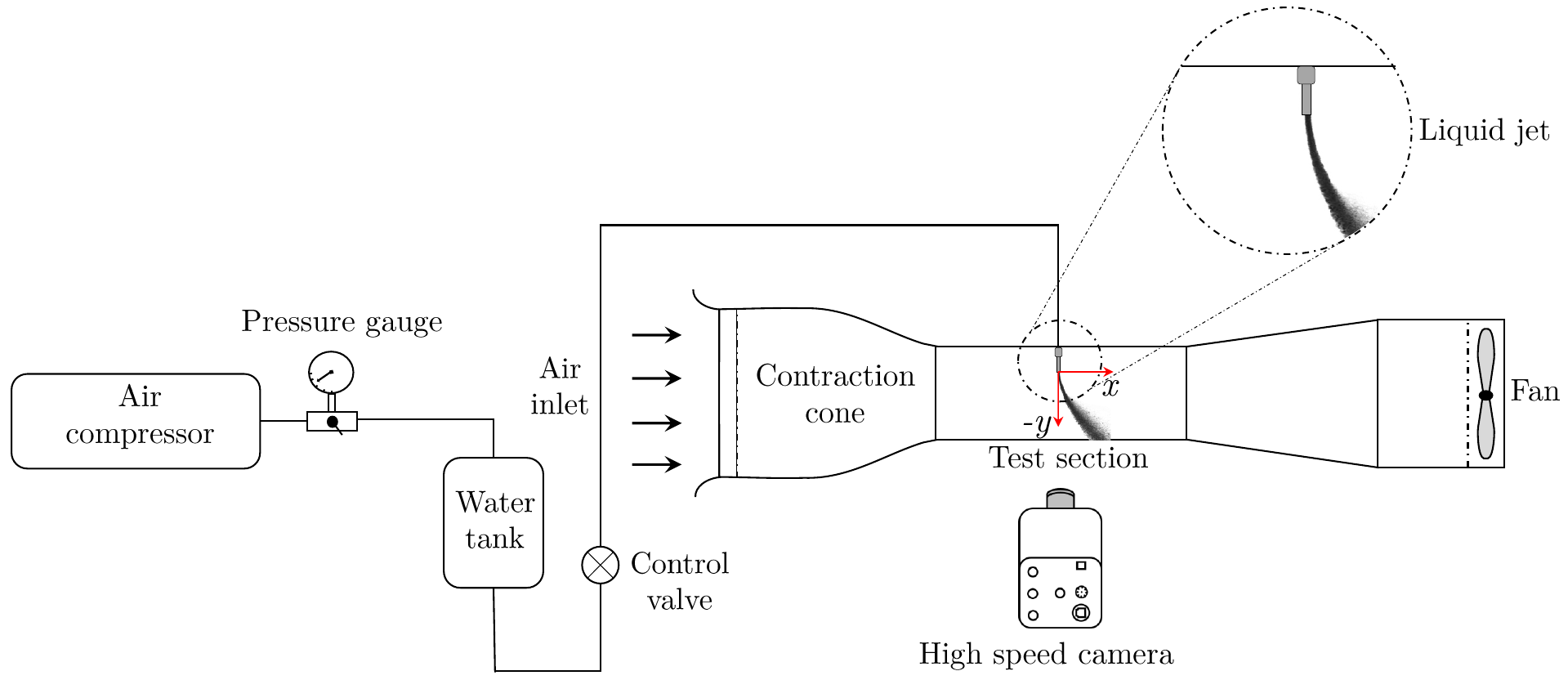}
\caption{A schematic of the experimental setup showing a liquid jet interacting with a crossflowing airstream inside an open-circuit, horizontal wind tunnel. The jet dispensing system consists of an air compressor, pressure regulator, water tank, control valve, and a flat-end 16-gauge needle. The origin of the coordinate system $(x,y,z)$ is defined at the needle tip, where $x$ denotes the airflow direction, negative $y$ corresponds to the jet direction, and $z$ represents the outward spanwise direction. The jet dynamics are captured using high-speed imaging and digital in-line holography.}
\label{fig:shadowgraphy_schematic}
\end{figure}

The dynamics of a continuous downward liquid jet in a crossflowing airstream is investigated using an open-circuit, horizontal subsonic wind tunnel facility, employing high-speed shadowgraphy and digital in-line holography (DIH). A schematic of the experimental setup is shown in figure \ref{fig:shadowgraphy_schematic}. The wind tunnel comprises a contraction section, an acrylic test section, a downstream diverging section, and a suction fan at the exit. Honeycombs and screens are installed at the entrance of the contraction section to straighten the incoming airflow. The facility generates a uniform airflow within the square cross-section test section of dimensions $300~\text{mm} \times 300~\text{mm} \times 600~\text{mm}$, driven by an axial fan located at the exit, whose speed is regulated using a variable-frequency drive (VFD), providing a volumetric flow rate in the range of $10^{-2}~\text{m}^3 \text{s}^{-1}$ to $3.5~\text{m}^3 \text{s}^{-1}$. In the present study, the average airstream velocity $(v_g)$ is varied from $0.5~\text{m} \text{s}^{-1}$ to $35~\text{m} \text{s}^{-1}$ and is measured using a Pitot–static tube mounted in the test section, connected to a calibrated differential pressure transducer and a data acquisition system. Prior to each experiment, the wind tunnel is operated for a sufficient duration to ensure that the flow attains a steady, fully developed state. 

The liquid jet is introduced through an injection mechanism located at the top-wall of the test section, as shown in figure \ref{fig:shadowgraphy_schematic}. The injection system comprises an air compressor, a precision pressure regulator, a pressurized tank filled to three-quarters capacity with deionized water, a ball valve for flow control, and a standard flat-end 16-gauge dispensing needle. The needle is mounted on the top wall of the test section and oriented such that the liquid jet is injected perpendicular to the wall and interacts with the airstream in a crossflow configuration, as illustrated in the inset of figure~\ref{fig:shadowgraphy_schematic}. The inner diameter of the needle is $d_{0} = 1.194~\text{mm}$. Since the liquid entry condition can influence the jet breakup process \citep{prakash2018liquid}, the nozzle geometry is kept identical throughout the study. The needle consists of a straight cylindrical internal passage of uniform diameter terminating in a flat exit, with no internal contraction, expansion, or swirl-generating features. The origin of the coordinate system $(x,y,z)$ is defined at the needle tip, where $x$ denotes the airflow direction, negative $y$ corresponds to the jet direction, and $z$ represents the outward spanwise direction. The pressurized reservoir ensures a steady and controlled liquid flow rate throughout the experiment. The volumetric flow rate of the liquid jet is determined by collecting the discharged liquid in a calibrated volumetric flask over a measured time interval. This procedure is repeated over a range of tank pressures to establish a reliable pressure-flowrate calibration curve, yielding a volumetric flow range of $5.25~\mu \text{L}~\text{s}^{-1}$ to $10.7~\mu\text{L}~\text{s}^{-1}$, corresponding to a jet velocity in the range $v_j = 4.7~\text{m} \text{s}^{-1}$–$9.6~\text{m} \text{s}^{-1}$. 

For the range of operating conditions considered in this study, the Froude number, defined as $Fr = v_g/\sqrt{g d_0}$, satisfies $Fr \gg 1$, indicating that the inertial forces associated with the crossflow are much larger than gravitational forces. Nevertheless, as discussed in the Introduction, gravity acts in the direction of the downward liquid jet and therefore influences the jet trajectory, droplet trajectories, residence time, and downstream evolution of the breakup process. Thus, although gravity plays a secondary role under the present operating conditions, its influence remains significant, while the breakup dynamics are governed predominantly by aerodynamic forcing. Using Buckingham’s $\pi$ theorem for the air–liquid system considered in the present study, the jet breakup dynamics in a crossflowing airstream can be characterized by two primary dimensionless parameters: the Weber number based on the airstream ($We_g \equiv {\rho_g v_g^2 d_0/\sigma}$) and the Weber number based on the liquid jet ($We_j \equiv {\rho_j v_j^2 d_0/\sigma}$). Here, $\rho_g = 1.225~\text{kg m}^{-3}$ and $\rho_j = 997~\text{kg m}^{-3}$ denote the densities of the gas and liquid, respectively, and $\sigma = 0.072~\text{N m}^{-1}$ represents the surface tension of the liquid. Additionally, as discussed in the introduction, several studies incorporate the momentum flux ratio, $Q \equiv \rho_j v_j^2 / \rho_g v_g^2 ~(\equiv We_j/We_g)$, as a dimensionless parameter representing the ratio of momentum flux between the liquid jet and the ambient crossflow. By independently varying the airstream velocity and the liquid jet velocity, the present study systematically examines the breakup dynamics over a wide parametric range, spanning gas Weber numbers $We_g = 0.5–21$ and liquid jet Weber numbers $We_j = 360–1500$.

\subsection{Shadowgraphy}\label{subsec:shadowgraphy}

\begin{figure}
\centering
\includegraphics[width=0.9\textwidth]{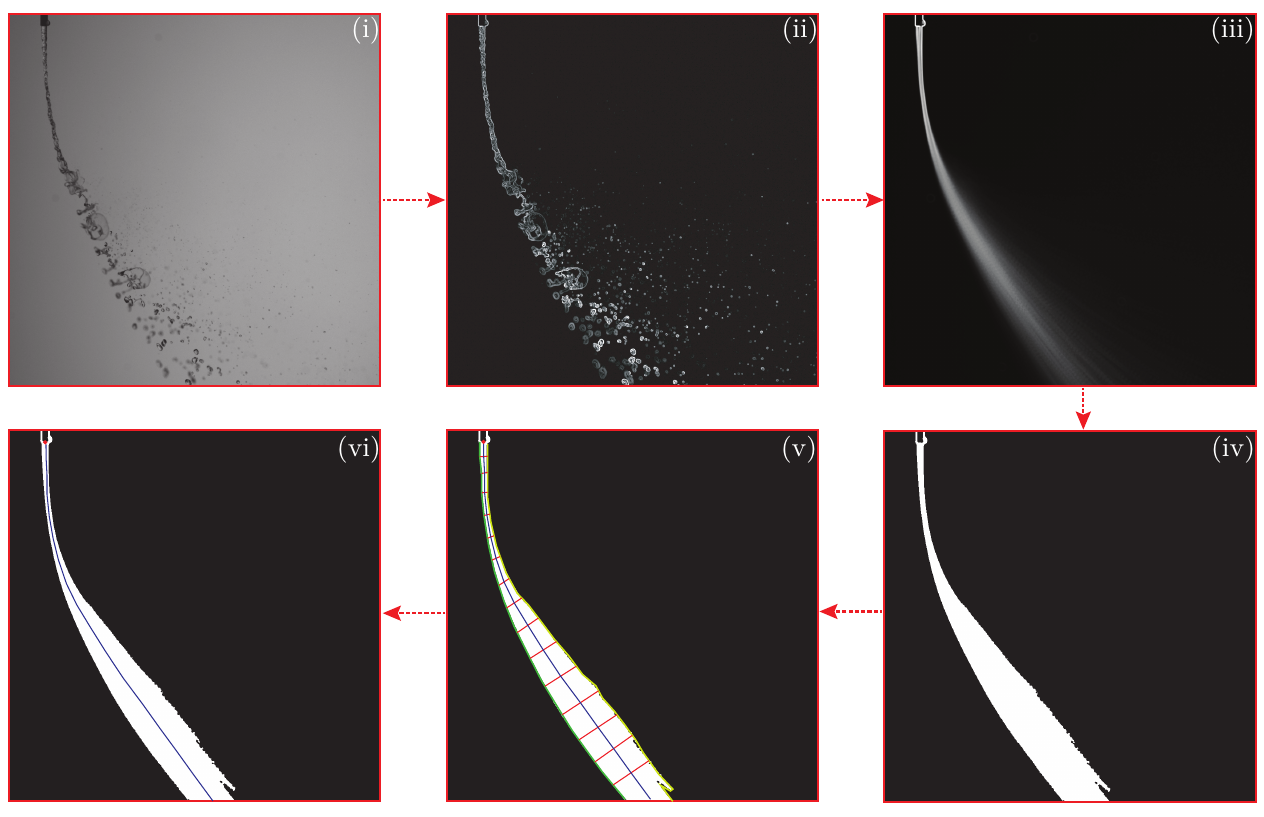}
\caption{Demonstration of the image-processing steps. (i) Original grayscale image; (ii) corresponding edge-detected image highlighting the jet boundaries; (iii) ensemble-averaged image obtained from 2000 edge-detected frames; (iv) binarised image using a thresholding method to isolate the jet region; (v) detection of jet boundaries and construction of normals between opposing edges; and (vi) extracted jet centreline trajectory.}
\label{fig:traj_image_processing}
\end{figure}

The characteristics of the jet are examined using shadowgraphy. The shadowgraphy system comprises a high-speed camera (Model: Photron Fastcam Nova S9; maximum resolution: $1024 \times 1024$ pixels) fitted with a macro lens (Model: Tokina atx-i 100 mm, $f$/2.8), a high-intensity LED light source, and a diffuser sheet to ensure spatially uniform background illumination. The images are acquired at 9000 frames per second (fps) with an exposure time of approximately $5~\mu$s using the high-speed camera. The field of view spans $106~\text{mm} \times 106~\text{mm}$, yielding a spatial resolution of $103~\mu\text{m}~\text{pixel}^{-1}$. A multi-step image processing workflow is applied to the shadowgraphy images to determine the trajectory and half-width of the jet. The complete sequence of steps is shown in figure~\ref{fig:traj_image_processing}. Grayscale images are first subjected to Canny edge detection to sharpen the jet boundaries. The edge-detected frames are then ensemble-averaged to produce a time-averaged intensity map representing the mean jet boundary. This map is subsequently binarised using an intensity thresholding criterion, and the outer boundary of the resulting binary image is extracted. The local jet half-width is computed by projecting normals from one boundary edge to the opposing edge, and the locus of the midpoints of these normals defines the jet centerline, thereby determining the jet trajectory.

\subsection{Digital in-line holography}\label{subsec:DIH}

\begin{figure}
\centering
\includegraphics[width=0.95\textwidth]{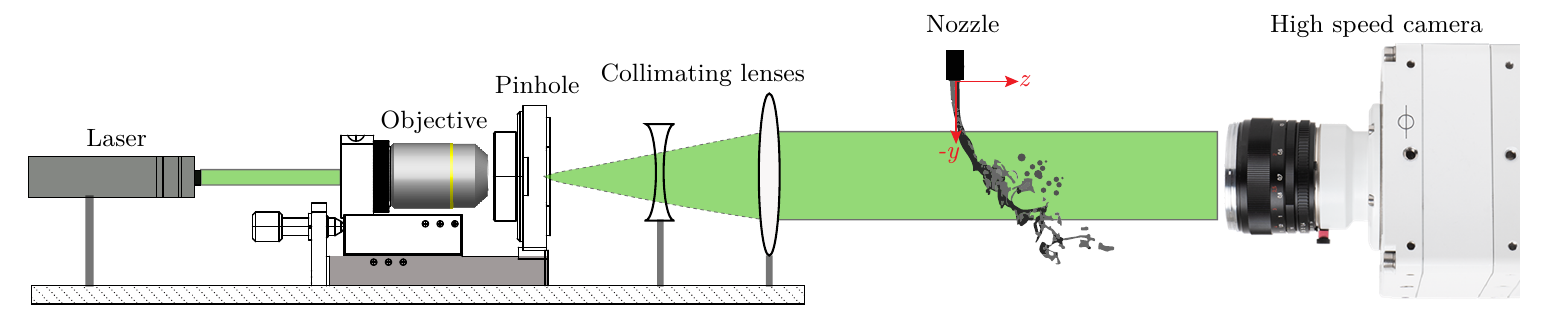}
\caption{Schematic (side view) of the digital in-line holography (DIH) setup used to measure the droplet size distribution resulting from liquid jet breakup in a crossflowing airstream.}
\label{fig:holography_schematic}
\end{figure}

The liquid jet, injected transversely into a high-speed airstream, undergoes rapid bending and deformation, followed by the growth of interfacial instabilities that culminate in breakup and atomization. The resulting volumetric droplet size distribution arising from jet fragmentation in the downstream region is analyzed using digital in-line holography (DIH). Figure~\ref{fig:holography_schematic} presents a schematic of the DIH system, with the optical axis aligned along the $z$-direction. Illumination is provided by a 100 mW, 532 nm continuous-wave laser (Model: SDL-532-100T, Shanghai Dream Lasers Technology Co.\ Ltd., China). The output beam is spatially filtered using an infinity-corrected plan-achromatic objective ($20\times$ magnification) and a $15~\mu$m pinhole (Holmarc Opto-Mechatronics Pvt.\ Ltd., India) to produce a clean, aberration-free wavefront. The filtered beam is then expanded using a bi-concave lens (diameter: 25 mm; focal length: $-50$ mm) and subsequently collimated by a bi-convex lens (diameter: 50 mm; focal length: 200 mm), yielding a spatially coherent beam of 50 mm diameter directed toward the region of interest. The interference patterns (holograms), formed by the superposition of the unscattered reference beam and the object-scattered wavefield, are recorded using a high-speed camera (Model: Photron Fastcam Nova S9) equipped with a 1:1 macro lens (Model: Mitakon, 200 mm focal length). The camera operates at 9000 fps with an exposure time of $2.5~\mu$s at a resolution of $1024 \times 1024$ pixels, yielding a spatial resolution of $20~\mu\text{m}~\text{pixel}^{-1}$.

The DIH measurement procedure comprises two main steps, namely hologram recording and numerical reconstruction. During the recording stage, the interference pattern formed between the object wave $E_{o}$ (laser light scattered by the droplets) and the reference wave $E_{r}$ (unscattered illumination) is captured on the camera sensor. The recorded intensity,
\begin{equation}
I_{H}(x,y) = \lvert E_{r} + E_{o}\rvert^{2}
           = \underbrace{\lvert E_{r}\rvert^{2}}_{\text{background}}
           + \lvert E_{o}\rvert^{2}
           + E_{r}^{*}E_{o} + E_{r}E_{o}^{*},
\label{eq:hologram_formation}
\end{equation}
therefore encodes both the amplitude and phase information of the scattered wavefield through the cross terms $E_{r}^{*}E_{o}$ and $E_{r}E_{o}^{*}$, where the asterisk denotes the complex conjugate. In the reconstruction stage, each hologram is numerically propagated with the reference beam to retrieve droplet information at different depths along the optical axis~\citep{katz2010applications}. 

Prior to numerical reconstruction, each hologram is pre-processed to remove background contributions. The background image is obtained by averaging a sequence of frames recorded without droplets in the test section and is subsequently subtracted from each raw hologram, suppressing the $\lvert E_{r}\rvert^{2}$ term in eq. \eqref{eq:hologram_formation} together with fixed-patternnoise and non-uniform illumination. Intensity normalization is then applied to equalise the intensity distribution and enhance droplet contrast, according to
\begin{equation}
I_{\mathrm{norm}}(x,y) = \frac{I - I_{\min}}{I_{\max} - I_{\min}},
\label{eqn:intensity}
\end{equation}
where $I_{\mathrm{norm}}$ denotes the normalised intensity, and $I_{\max}$ and $I_{\min}$ represent the maximum and minimum intensities of the hologram, respectively.

Following pre-processing, numerical reconstruction is performed using the Rayleigh–Sommerfeld diffraction formulation, which can be expressed in the spatial domain as a convolution. This is given by
\begin{equation}
I_{r}(x,y,z) = I_{h}(x,y) \otimes h(x,y,z),
\label{eqn:recon_intensity}
\end{equation}
where $I_{r}(x,y,z)$ denotes the intensity of the reconstructed wavefield, $I_{h}(x,y)$ is the pre-processed hologram, $\otimes$ represents the convolution operation, and $h(x,y,z)$ is the Rayleigh-Sommerfeld diffraction kernel. For computational efficiency, this convolution is carried out in the Fourier domain, whose free-space transfer function over a propagation distance $z$ is given by
\begin{equation}
H(f_{x},f_{y},z)=\exp\!\left(jkz\sqrt{1-\lambda^{2}f_{x}^{2}-\lambda^{2}f_{y}^{2}}\,\right).
\label{eq:as_kernel}
\end{equation}
Here $\lambda$ is the illumination wavelength, $k=2\pi/\lambda$ the associated wavenumber, $j=\sqrt{-1}$, and $(f_{x},f_{y})$ the transverse spatial-frequency coordinates conjugate to $(x,y)$. Propagation to any chosen depth thus reduces to a spectral product; the hologram is Fourier transformed, multiplied by \eqref{eq:as_kernel}, and inverse transformed back to real space,
\begin{equation}
I_{r}(x,y,z)=\mathrm{FFT}^{-1}\!\left\{\mathrm{FFT}\!\left[I_{h}(x,y)\right]\times H(f_{x},f_{y},z)\right\},
\label{eq:as_reconstruction}
\end{equation}
with $\mathrm{FFT}$ and $\mathrm{FFT}^{-1}$ the forward and inverse fast Fourier transforms. The in-focus intensity at each plane is recovered from the magnitude of the complex field $\lvert I_{r}(x,y,z)\rvert$. The reconstruction volume spans a depth of 140~mm, covering the entire region of interest, with adjacent reconstructed planes separated by $\Delta z = 40~\mu\text{m}$.

\begin{figure}
\centering
\includegraphics[width=1.0\textwidth]{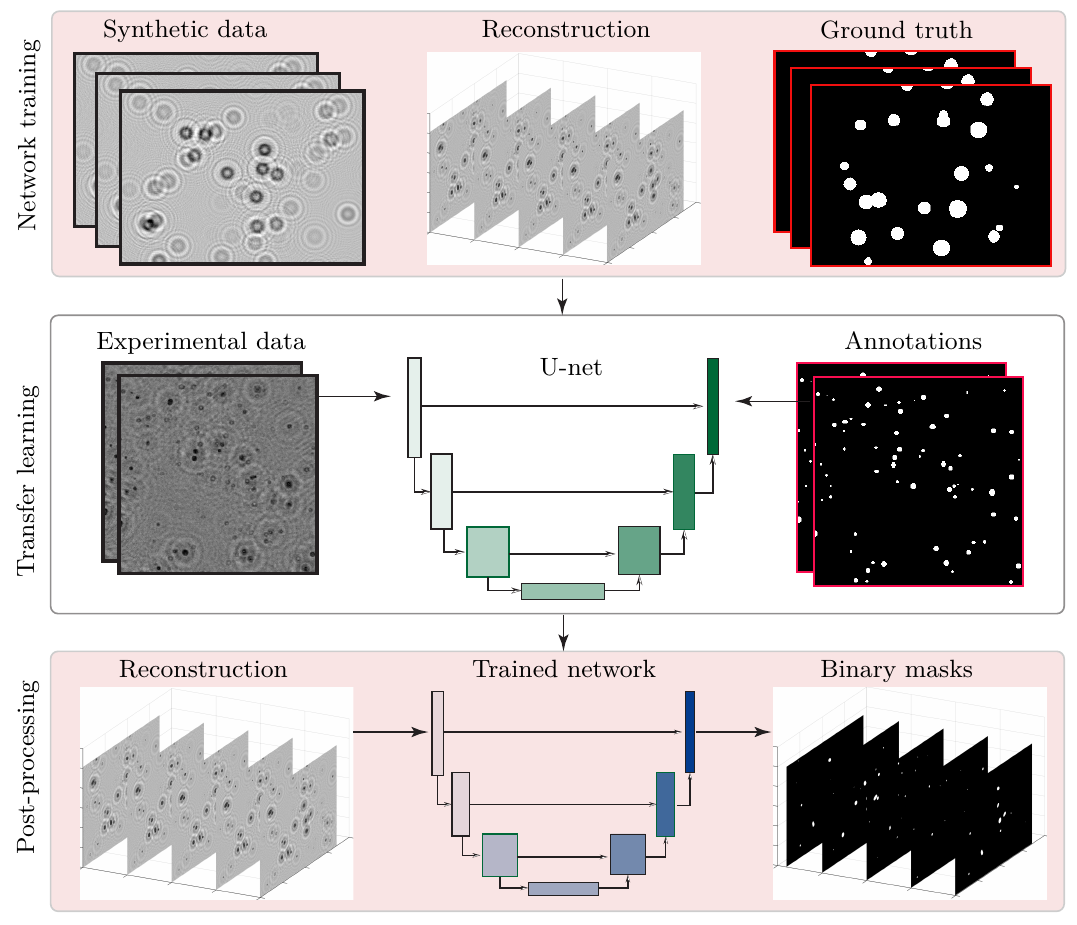}
\caption{Different steps employed in the digital in-line holography (DIH) methodology, illustrating the U-Net training procedure and subsequent post-processing of reconstructed experimental holograms.}
\label{fig:machine_learning}
\end{figure}

The reconstructed volume contains droplets distributed across multiple depth planes. Accurate identification of droplet boundaries and size estimation is challenging using conventional image processing techniques due to the coexistence of sharply focused particles and out-of-focus fringe patterns in the reconstructed holograms. To address this, a machine-learning-based semantic segmentation approach is adopted, following the framework of \cite{ade2022droplet,ade2024application} with enhancements through transfer learning. The method employs a U-Net architecture~\citep{Ronneberger2015}, and the overall workflow is illustrated in figure~\ref{fig:machine_learning}. The workflow comprises three sequential stages: (i) network training on synthetic data, (ii) transfer learning on experimental images, and (iii) post-processing of the reconstructed holograms. The U-Net is a fully convolutional network with a symmetric encoder-decoder structure with an input image size of 256$\times$256. The encoder alternates convolutional blocks with pooling operations; each pooling step halves the spatial resolution, broadening the effective receptive field so that progressively more global context is captured as the feature representation is compressed. The decoder reverses this process, progressively restoring the full spatial resolution via learned transposed convolution upsampling. At each scale, lateral skip connections reintroduce the corresponding high-resolution feature maps from the encoder, thereby preserving fine-scale spatial information and sharp droplet boundaries that might otherwise be lost during downsampling \citep{Ronneberger2015}. The hidden units employ a leaky rectified linear unit (ReLU) activation function,
\begin{equation}
f(\zeta)=
\begin{cases}
\zeta, & \zeta \geq 0,\\[2pt]
0.01\,\zeta, & \zeta <0,
\end{cases}
\label{eq:leaky_relu}
\end{equation}
in which $\zeta$ is the pre-activation (the weighted sum of inputs at a unit). The small negative slope preserves a non-zero gradient on the negative branch, thereby avoiding the inactive-unit behaviour to which the plain rectifier is prone. Because segmentation is posed here as per-pixel binary labelling (droplet versus background), the output layer applies a sigmoid, $\sigma(\zeta)=1/(1+e^{-\zeta})$, returning for every pixel the probability that it lies on a droplet. Since obtaining a large set of annotated experimental images is impractical, a dataset of 10,000 synthetic holograms of spherical particles spanning a wide range of sizes is first generated using numerical simulations following the work of \cite{Latychevskaia:15}. The simulated set is further diversified through geometric augmentations, including random shifts, rotations, and shear transforms applied to the reconstructed images, which expose the network to the variability in droplet positions, orientations, and apparent distortions found in real recordings \citep{frid2018synthetic}. The reconstructed synthetic holograms, along with their corresponding ground-truth annotations, are used to train the U-Net and obtain a base model. The network is then trained on an NVIDIA RTX 4090 Ti GPU for 1000 epochs, using a batch size of 32 and a learning rate of $10^{-5}$. The input layer is fixed at a resolution of $256 \times 256$ pixels. To process larger images that exceed this input size, a patch-based inference strategy is employed. Specifically, input images are divided into patches of size $256 \times 256$ pixels with a 32-pixel overlap. The predictions from individual patches are then stitched back together to reconstruct the full-resolution output.

This base model is subsequently refined through transfer learning using 150 manually annotated experimental images extracted from the reconstructed volume. Ground-truth masks are generated by applying local thresholding around individual droplets in the reconstructed holograms. To ensure robustness and general applicability, experimental images from multiple flow conditions are included in the transfer learning dataset. The trained network produces binary segmentation masks of droplets at each reconstructed depth plane. These binary images are then processed in \textsc{Matlab}$^{\circledR}$ using the \textsc{regionprops} function. Morphological filtering based on area, circularity, and solidity is applied to eliminate noise, partial detections, and non-droplet structures. Additionally, droplets with diameters smaller than two pixels are excluded from the analysis. The droplet diameters obtained from all datasets are finally aggregated to construct a comprehensive size distribution. For each set of experimental conditions, the droplet size distributions are obtained by averaging over three independent repetitions. Furthermore, to minimize uncertainties associated with liquid flow-rate control, crossflow-velocity measurements, and image resolution, the droplet statistics are averaged over approximately 100 independent realizations, comprising a total of approximately 142,000 reconstructed droplets across all operating conditions. This procedure ensures good statistical convergence, improves measurement repeatability, and enhances the robustness of the observed trends. 

\begin{figure}
\centering
\includegraphics[width=0.95\textwidth]{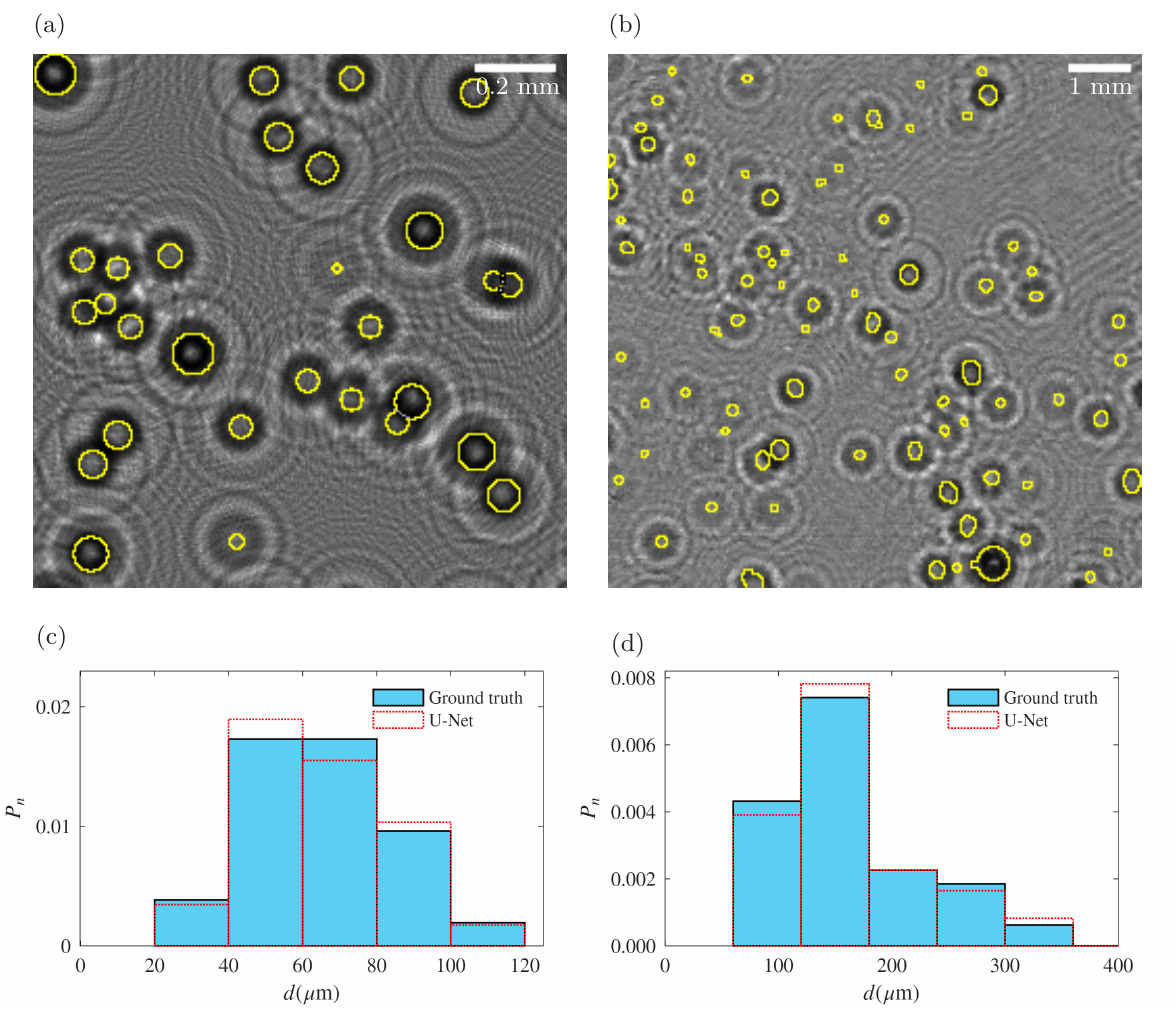}
\caption{Validation of droplet size distributions using U-Net-based holography. (a) Synthetic hologram with overlaid U-Net-predicted droplet boundaries. (b) Experimental hologram of a jet in cross-flow with overlaid U-Net-predicted droplet boundaries. Comparison of the number probability distribution, $P_n$, obtained from U-Net-based holography with manual annotations (ground truth) for (c) the synthetic hologram shown in panel (a), and (d) the experimental hologram shown in panel (b).}
\label{fig:holography_validations}
\end{figure}

The accuracy and reliability of the digital inline holography (DIH) reconstruction and droplet-detection methodology are validated using both synthetic and experimental holograms. To assess the accuracy of droplet sizing based on the U-Net framework, the predicted droplet boundaries are compared with the simulated masks for the synthetic holograms and with manually annotated ground-truth boundaries for the experimental holograms. Figure~\ref{fig:holography_validations}(a) shows a representative synthetic hologram with the U-Net-predicted droplet boundaries overlaid, while figure~\ref{fig:holography_validations}(b) presents a typical experimental hologram of a jet in cross-flow, similarly annotated with the predicted boundaries. Qualitatively, the predicted boundaries closely follow the droplet contours in both cases, capturing the range of droplet sizes within the field of view. For a quantitative assessment, the number probability distribution ($P_n$) calculated using equation~\eqref{npd}, is compared between the U-Net predictions and the corresponding ground-truth annotations for the synthetic and experimental holograms (figures~\ref{fig:holography_validations}c and \ref{fig:holography_validations}d, respectively). The predicted distributions show close agreement with the ground truth across the range of droplet sizes, with a median error of less than 2\% and a mean error of less than 1\%.

\section{Results and discussion} \label{sec:dis}

We begin the presentation of our results by examining the distinct breakup dynamics and resulting morphology of a continuous downward liquid jet in a crossflowing airstream in terms of the governing dimensionless parameters, the gas Weber number ($We_g$), the liquid Weber number ($We_j$), and the momentum flux ratio ($Q$). These parameters collectively govern the onset of instabilities, the evolution of breakup modes (jet morphology), and the spatial variation of droplet size distributions.

\subsection{Morphology} \label{sec:morphology}

\begin{figure}
\centering
(a) \hspace{2.4cm} (b) \hspace{2.4cm} (c) \hspace{2.4cm} (d) \\
\includegraphics[width=0.9\textwidth]{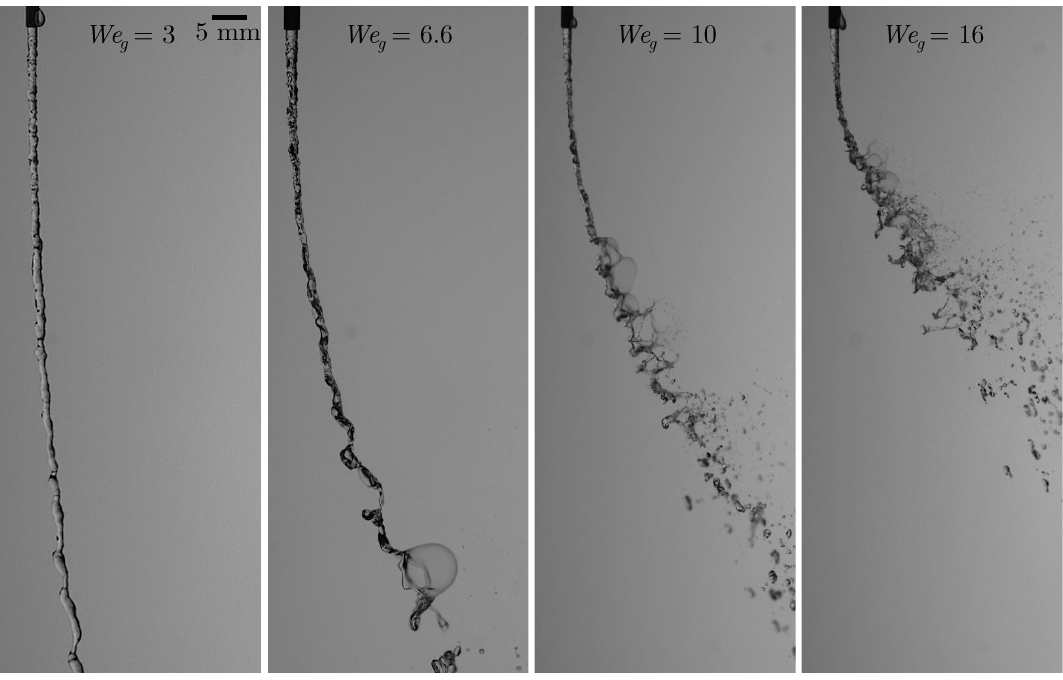}
\caption{Dynamics and morphology of a continuous downward liquid jet in a crossflowing airstream exhibiting distinct fragmentation behaviors for different values of the airstream Weber number ($We_g$) at a fixed liquid jet Weber number of $We_j = 1050$.}
\label{fig:traj_we_j_con_image}
\end{figure}

Figure \ref{fig:traj_we_j_con_image} illustrates the evolution of jet breakup dynamics and morphology with increasing airstream Weber number ($We_g$) at a fixed liquid Weber number ($We_j = 1050$), thereby isolating the role of aerodynamic forcing in the fragmentation process. At the lowest gas Weber number considered, $We_g = 3$ (figure \ref{fig:traj_we_j_con_image}a), the jet maintains a long, coherent liquid column prior to breakup. The penetration depth is  large, and the jet undergoes only mild bending under the crossflow of the airstream, indicating that liquid inertia dominates over aerodynamic drag. In this regime, breakup occurs far downstream through a columnar disintegration mediated by surface undulations. This column-type jet breakup is governed by the Rayleigh-Plateau (RP) instability. As $We_g$ increases to $We_g = 6.6$ (figure~\ref{fig:traj_we_j_con_image}b), aerodynamic drag becomes more pronounced, leading to enhanced jet deflection and a corresponding reduction in penetration depth, defined as the axial distance from the needle exit to the onset of bag formation or disintegration. The stronger deceleration on the windward side promotes Rayleigh–Taylor (RT) instability, which governs the primary mode of disintegration in this regime. Under this moderate aerodynamic loading, the crossflow remains insufficient to induce significant Kelvin–Helmholtz (KH) type surface stripping. Instead, aerodynamic forcing deforms and flattens the liquid column, progressively inflating it into thin, balloon-like bag structures. These bags subsequently rupture due to the growth of RT instabilities on the liquid film. Meanwhile, the relatively high liquid inertia sustains thicker rim structures and ligaments, which undergo further breakup via Rayleigh–Plateau instability, producing larger secondary droplets. At $We_g = 10$ (figure \ref{fig:traj_we_j_con_image}c), the jet undergoes substantial deflection in the streamwise direction, with a pronounced curvature and a much shorter intact core. Breakup initiates close to the injection point, with Rayleigh–Taylor instability dominating the formation and inflation of multiple bags along the jet column. This multibag or interacting-bag regime is characterized by rapid inflation and rupture of liquid membranes, accompanied by intense ligament formation and secondary atomization. Consequently, instability growth is amplified, surface stripping is enhanced, and a dense, widely dispersed droplet field is produced. As we increase the gas Weber further ($We_g = 16$; figure \ref{fig:traj_we_j_con_image}d), the jet undergoes severe deformation and strong deflection, becoming rapidly aligned with the crossflow direction. The intact core length is minimal, indicating near-immediate disintegration after injection. As a result, the breakup process involves a combination of multibag and shear-driven mechanisms, where Rayleigh–Taylor instability continues to drive large-scale deformation and bag fragmentation, while strong interfacial shear may additionally contribute to aerodynamic stripping through the development of Kelvin–Helmholtz-type surface disturbances. Thus, the breakup process involves a combination of multibag and shear-driven mechanisms, with intense aerodynamic forcing promoting the formation of highly unstable bag structures and surface stripping. As a result, the liquid column rapidly fragments into fine ligaments and droplets, producing a highly dispersed spray with predominantly small-scale and unsteady structures.

\begin{figure}
\centering
(a) \hspace{2.4cm} (b) \hspace{2.4cm} (c) \hspace{2.4cm} (d) \\
\includegraphics[width=0.9\textwidth]{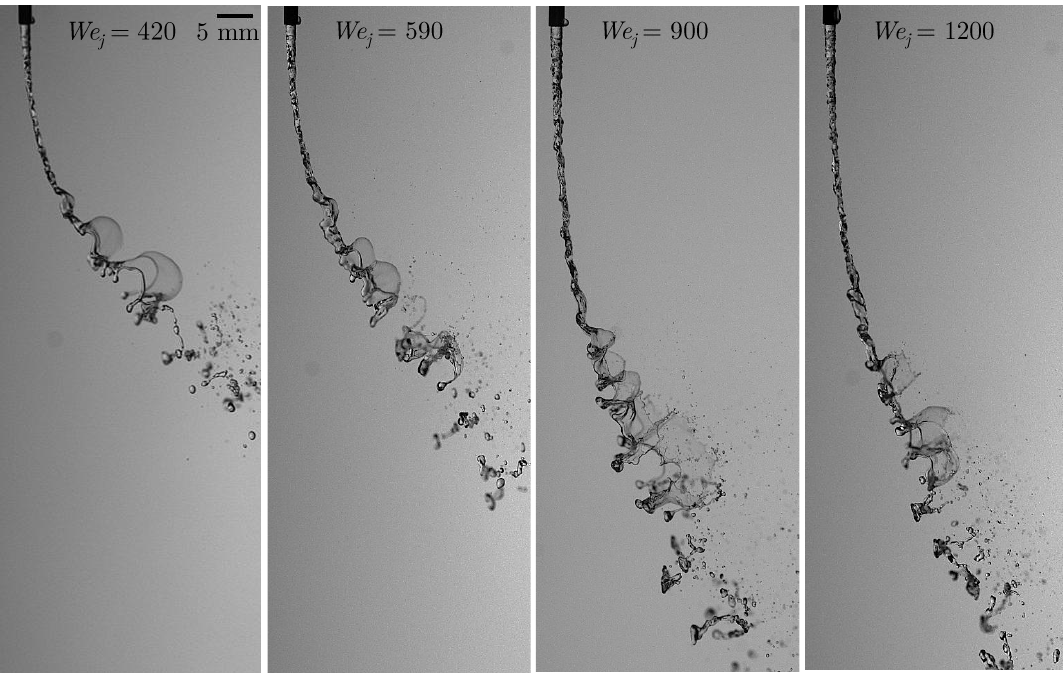}
\caption{Dynamics and morphology of a continuous downward liquid jet in a crossflowing airstream exhibiting distinct fragmentation behaviors for different values of the liquid jet Weber number ($We_j$) at a fixed airstream Weber number of $We_g = 8.17$.}
\label{fig:traj_we_g_con_image}
\end{figure}

Next, focusing on the bag breakup regime, figure \ref{fig:traj_we_g_con_image} examines the effect of the liquid Weber number ($We_j$) on the evolution of jet breakup dynamics and morphology at a fixed airstream Weber number ($We_g = 8.17$). This enables us to analyze the role of liquid inertia in the fragmentation process under a fixed airstream condition. At $We_j = 420$ (figure \ref{fig:traj_we_g_con_image}a), the jet exhibits limited penetration into the crossflow and undergoes significant bending, indicating the dominance of aerodynamic drag over liquid inertia. The intact core is short, and bag formation occurs close to the nozzle. The flow is characterized by isolated bags, primarily driven by RT instability on the windward side. As $We_j$ increases to 590 (figure \ref{fig:traj_we_g_con_image}b), the jet penetrates further and the trajectory becomes less curved, reflecting the increasing influence of liquid inertia. The intact core extends downstream, and the breakup structures become more pronounced, with emerging interactions between neighboring bags. At $We_j = 900$ (figure \ref{fig:traj_we_g_con_image}c), the jet shows significantly greater penetration and adopts a more vertical trajectory. Although the intact core length increases, the onset of breakup shifts slightly downward due to higher jet momentum. Multiple bags and nodes develop along the jet column, indicating a transition to an interacting-bag regime dominated by RT instability. At the highest value of the jet Weber number considered, $We_j = 1200$ (figure \ref{fig:traj_we_g_con_image}d), liquid inertia strongly dominates, resulting in deep penetration and minimal deflection by the crossflow. Despite a longer intact core, the breakup becomes more vigorous, with rapid formation and disintegration of bags, rims, and ligaments. The combined effects of strong Rayleigh–Taylor and Rayleigh–Plateau instabilities, together with shear-induced Kelvin–Helmholtz waves, enhance secondary atomization and produce a dense, widely dispersed droplet field.

\begin{figure}
\centering
\includegraphics[width=0.58\textwidth]{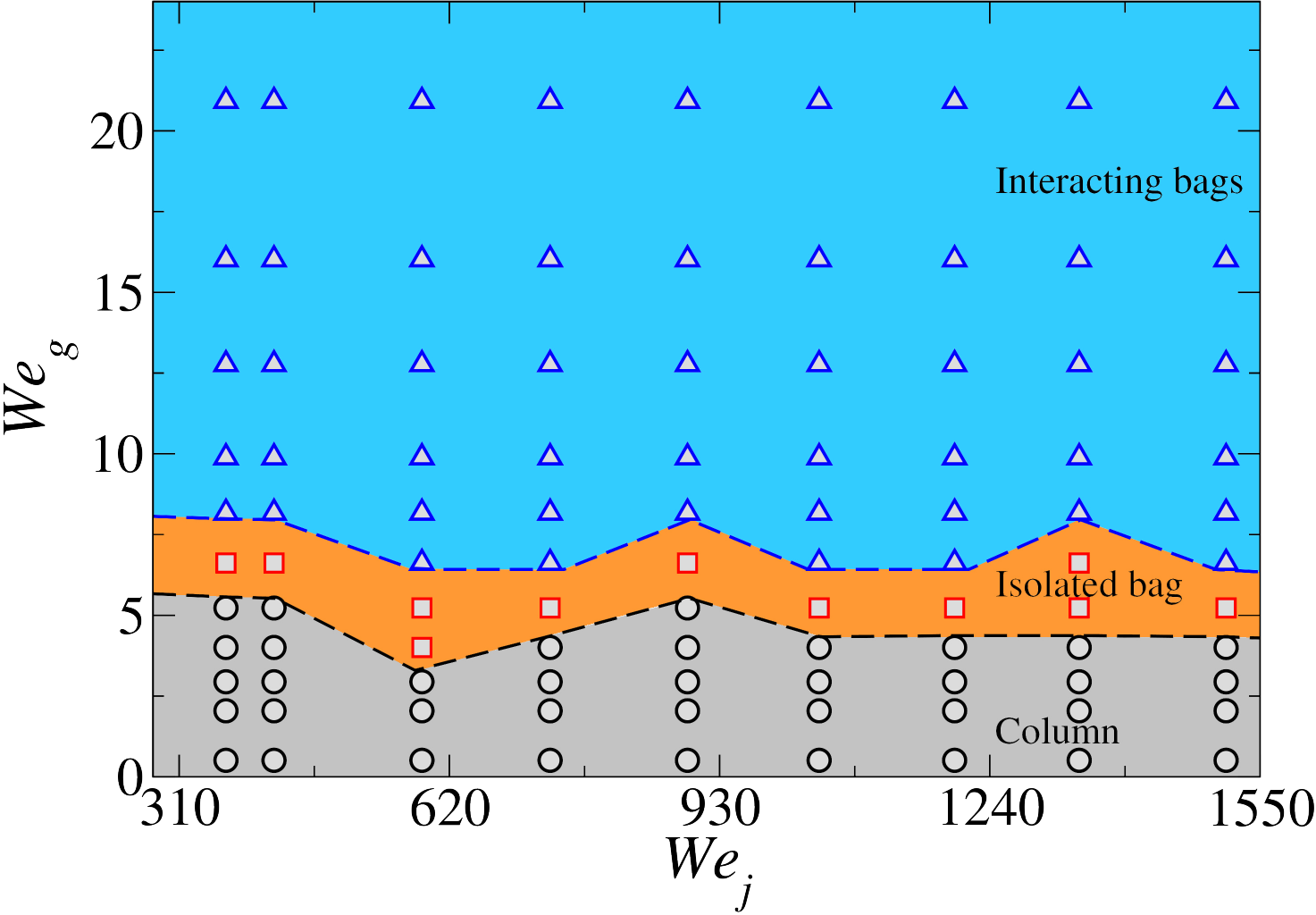}
\caption{Regime map demarcating different jet breakup dynamics, namely column breakup (circles), isolated bag breakup (squares), and interacting bags (triangles), in the $\We_g$–$\We_j$ space.}
\label{fig:regime_map}
\end{figure}

Overall, figures \ref{fig:traj_we_j_con_image} and \ref{fig:traj_we_g_con_image} demonstrate that increasing $We_g$ drives transitions from column to bag, multibag, and interacting-bag breakup, while increasing $We_j$ primarily enhances jet penetration and influences the spatial evolution of fragmentation process. This motivates the construction of a phase diagram delineating the different breakup regimes, as shown in figure \ref{fig:regime_map} in the $We_g$–$We_j$ parameter space. The isolated- and interacting-bag cases may be regarded as sub-regimes of the broader bag-breakup regime. The distinction is introduced to emphasize the different interaction dynamics observed during the breakup process. In the isolated-bag regime, a bag evolves and ruptures without interacting with neighbouring bags, whereas in the interacting-bag regime, adjacent bags come into contact and their liquid rims interact prior to breakup, thereby modifying the subsequent breakup dynamics. It should be noted that the present classification is based on observations within the field of view ($106~\mathrm{mm} \times 106~\mathrm{mm}$). Accordingly, the term isolated bag refers to a bag that evolves without interaction with other bags within the observed field of view, although additional bags may exist outside the imaging region.

It can be observed that at low gas Weber numbers ($We_g < 5$), the jet undergoes column breakup over the entire range of $We_j$ considered in the present study ($We_j<1550$), indicating that aerodynamic forcing is insufficient to induce significant large-scale deformation. In this regime, breakup is governed primarily by RP instability and surface-wave-driven disintegration. As $We_g$ increases within the range $5 < We_g < 6.5$, the flow transitions to isolated bag breakup, characterized by the inflation of the liquid surface under enhanced aerodynamic loading. In this regime, the breakup is primarily governed by Rayleigh–Taylor instability, while the relatively high liquid inertia sustains thicker rim structures and ligaments that subsequently fragment through Rayleigh–Plateau instability, leading to the formation of larger secondary droplets. For $We_g \approx 8$, the flow enters a multibag or interacting-bag regime across all $We_j$, where strong aerodynamic stresses generate multiple, simultaneously evolving bag structures and promote intense atomization. The breakup process is characterized by a combined multibag mechanism, in which Rayleigh–Taylor instability remains the dominant driver of large-scale deformation and bag fragmentation. Within the range of operating conditions investigated in the present study, the observed regime transitions correlate predominantly with the gas Weber number ($We_g$). It should be noted, however, that previous studies have shown that liquid-side parameters, including $Re_j$, $We_j$, and injector geometry, can also influence the breakup behaviour and regime transitions \citep{sinha2023effect}. Although the regime boundaries shown in figure \ref{fig:regime_map} are approximate, they clearly demonstrate that the gas Weber number primarily governs the transition between breakup modes, whereas the liquid Weber number mainly influences the penetration depth and spatial evolution of fragmentation. These effects are discussed in the following section and form the basis for establishing the theoretical framework for the droplet size distribution presented subsequently.

\subsection{Jet characteristics} \label{sec:characteristics}

\begin{figure}
\centering
 \hspace{0.6cm} (a) \hspace{6.2cm} (b) \\
\includegraphics[width=0.45\textwidth]{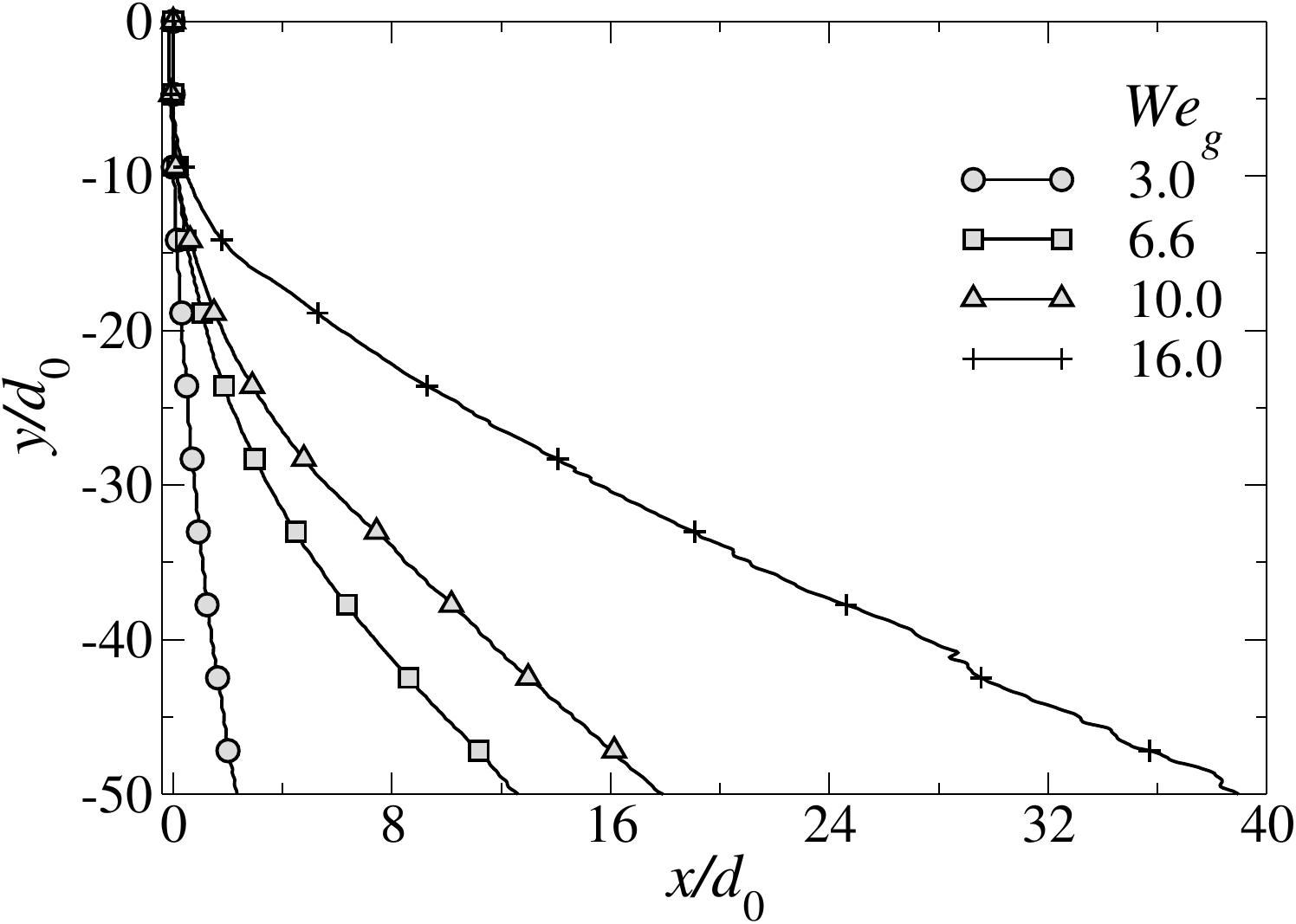} \hspace{5mm}
\includegraphics[width=0.45\textwidth]{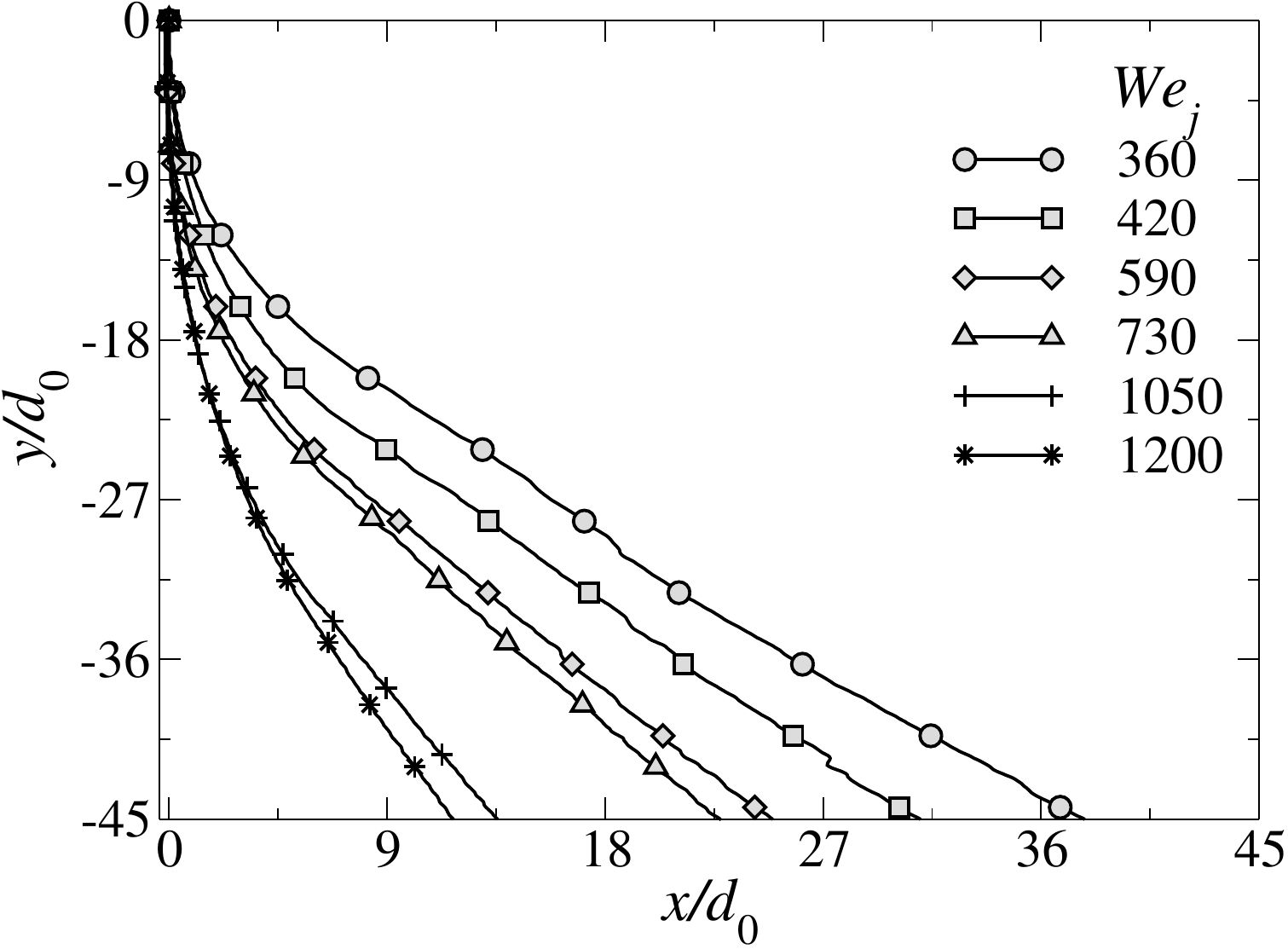}
\caption{Trajectory of the jet for different values of (a) $We_g$ at $We_j = 1050$ and (b) $We_j$ at $We_g = 8.17$, presented in the normalized plane $(x/d_0, ~ y/d_0)$.}
\label{fig:jet_trajectory}
\end{figure}

\begin{figure}
\centering
 \hspace{0.6cm} (a) \hspace{6.2cm} (b) \\
\includegraphics[width=0.45\textwidth]{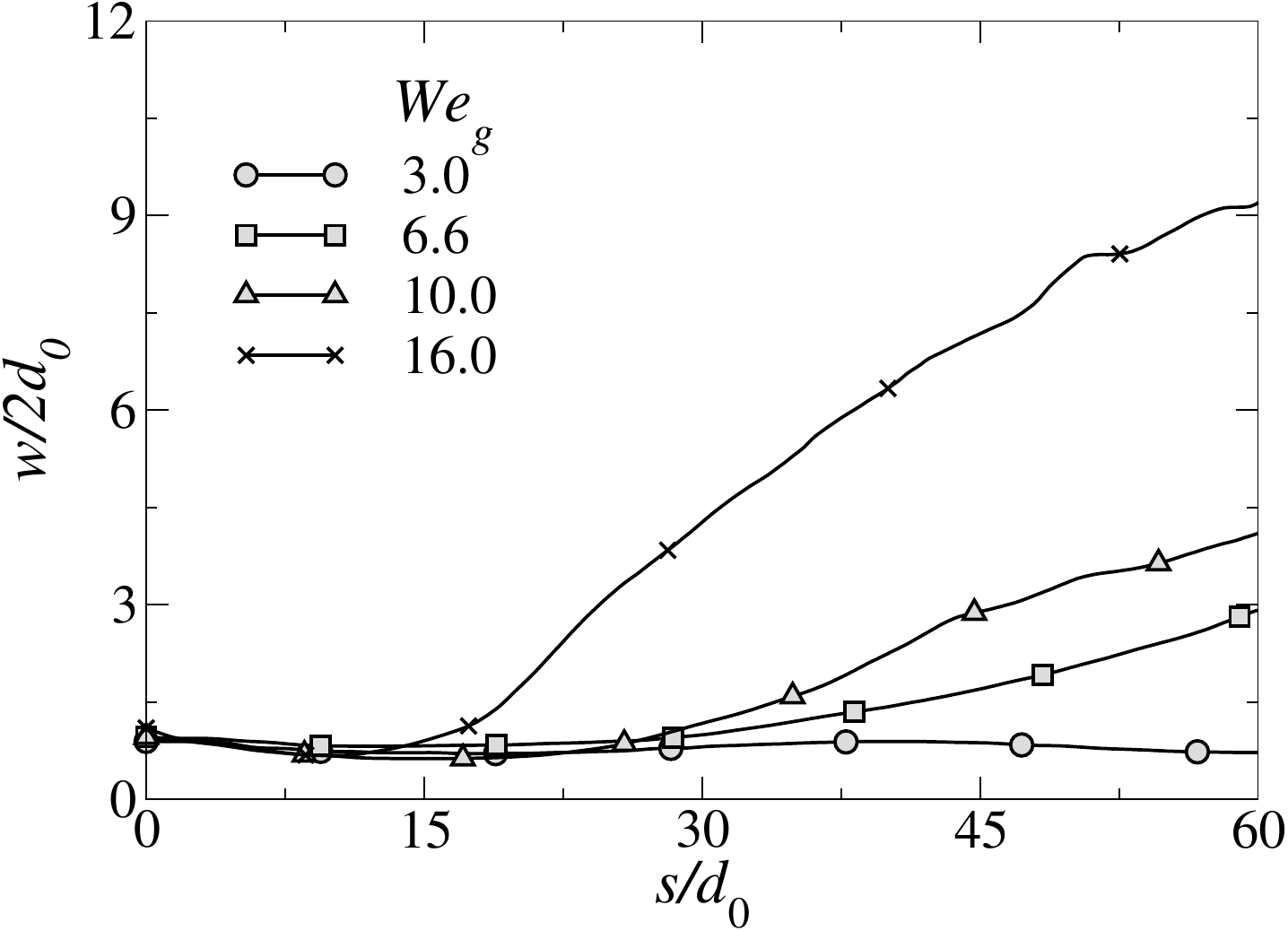}
\includegraphics[width=0.45\textwidth]{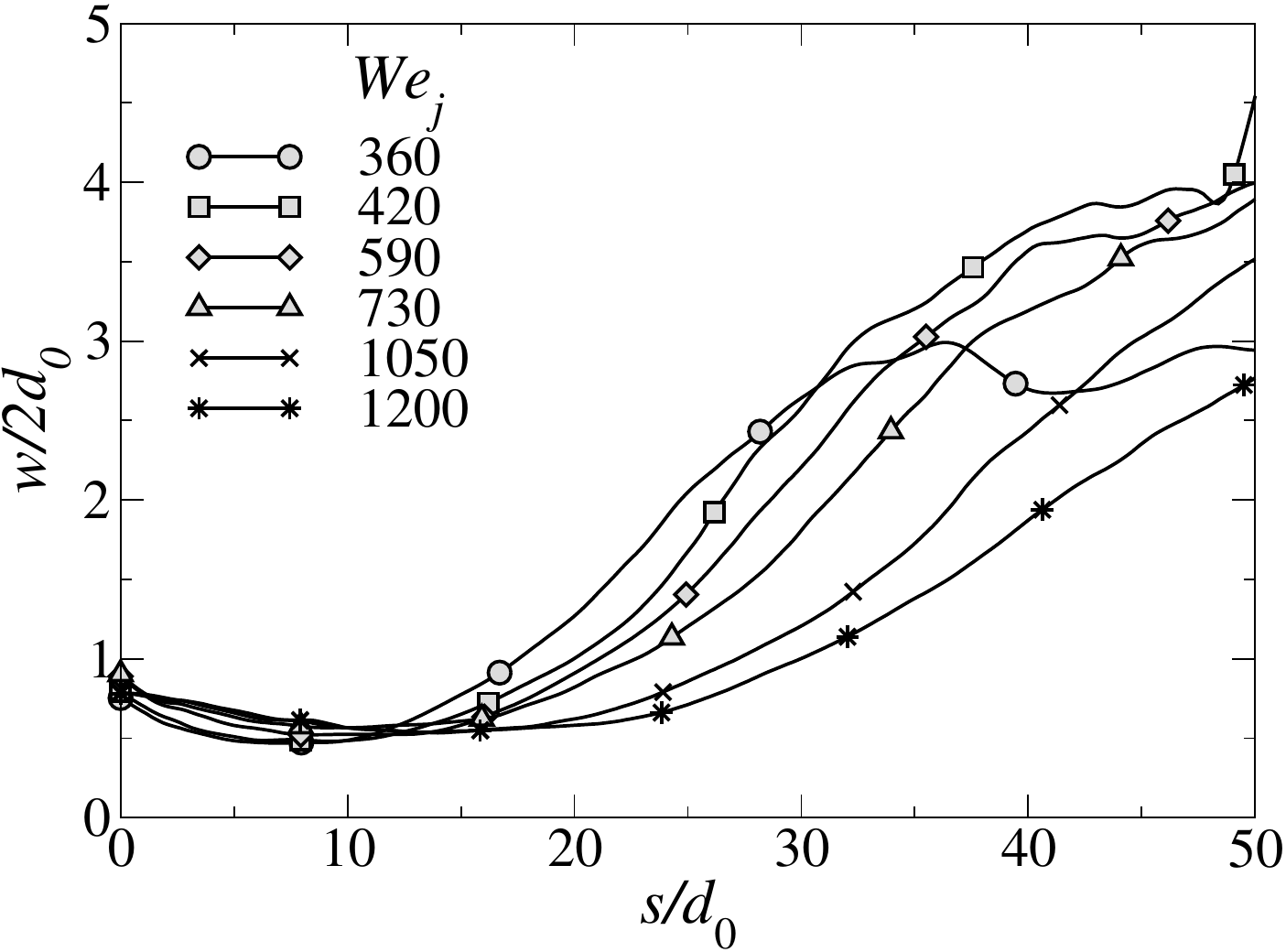}
\caption{Variation of the normalized half-width ($w/2d_0$) along the curved trajectory of the liquid jet ($s/d_0$) for different values of (a) $We_j$ at $We_g = 8.17$, and (b) $We_g$ at $We_j = 1050$.}
\label{fig:spread_width}
\end{figure}

Next, we analyze the geometric characteristics of a jet transversely injected into a crossflowing airstream for different values of $We_g$ and $We_j$, using the image-processing technique presented in figure~\ref{fig:traj_image_processing}. Figures \ref{fig:jet_trajectory} and \ref{fig:spread_width} quantify the geometric evolution of the jet in terms of its trajectory and lateral spreading, providing further insight into the effects of $We_g$ and $We_j$. The jet trajectories for different values of $We_g$ at $We_j = 1050$ and for different values of $We_j$ at $We_g = 8.17$, plotted in the normalized plane $(x/d_0,~ y/d_0)$, are shown in figure~\ref{fig:traj_we_j_con_image}(a) and \ref{fig:traj_we_j_con_image}(b), respectively. As shown in figure~\ref{fig:traj_we_j_con_image}(a), at fixed $We_j$, increasing $We_g$ enhances aerodynamic forcing, leading to stronger jet bending and reduced penetration. This behavior is consistent with the earlier observed transition from column to bag and multibag breakup regimes. The increased deflection reflects the growing dominance of aerodynamic stresses, which also promote the onset of Rayleigh–Taylor instability and early fragmentation. Conversely, at fixed $We_g = 8.17$ (figure~\ref{fig:traj_we_j_con_image}b), increasing $We_j$ results in deeper penetration and reduced jet deflection, as higher liquid inertia enables the jet to resist aerodynamic drag better. This behavior is also consistent with the earlier observation that increasing $We_j$ primarily alters the global jet structure without fundamentally changing the breakup regime. In particular, the more vertical jet trajectories at higher $We_j$ correspond to longer intact cores and delayed lateral dispersion, even though the flow remains within the bag or interacting-bag regime governed by $We_g$.

Figure \ref{fig:spread_width} shows the evolution of the normalized jet half-width along the jet trajectory. At a fixed $We_j$ (figure \ref{fig:spread_width}a), an increase in $We_g$ leads to a substantial increase in the jet half-width, indicating enhanced deformation and lateral spreading driven by stronger aerodynamic loading. This trend is consistent with the transition toward multibag and interacting-bag breakup regimes, in which Rayleigh–Taylor instability promotes significant transverse growth of the jet structure. In contrast, at constant $We_g$ (figure \ref{fig:spread_width}b), the jet half-width decreases with increasing $We_j$, suggesting that higher liquid inertia inhibits lateral expansion and preserves jet coherence over longer streamwise distances. This observation is consistent with the delayed onset of fragmentation at higher $We_j$, even though instability growth persists. Overall, these results highlight that $We_g$ primarily dictates the breakup mode and atomization intensity, whereas $We_j$ controls jet penetration and the spatial development of the flow. Consequently, the jet geometry, including its trajectory, bending, and spreading, remains strongly coupled to the underlying instability dynamics, providing a clear macroscopic signature of the distinct breakup regimes identified in the regime map. Furthermore, identifying regions along the jet where the droplet size distribution changes significantly helps determine the appropriate spatial locations for detailed holographic measurements, discussed in Section \ref{subsec:holography_dsd}

\subsection{Theoretical modeling of jet morphology and comparison with experiments} \label{sec:theory}

\begin{figure}
\centering
 \hspace{0.6cm} (a) \hspace{5.8cm} (b) \\
\includegraphics[width=0.45\textwidth]{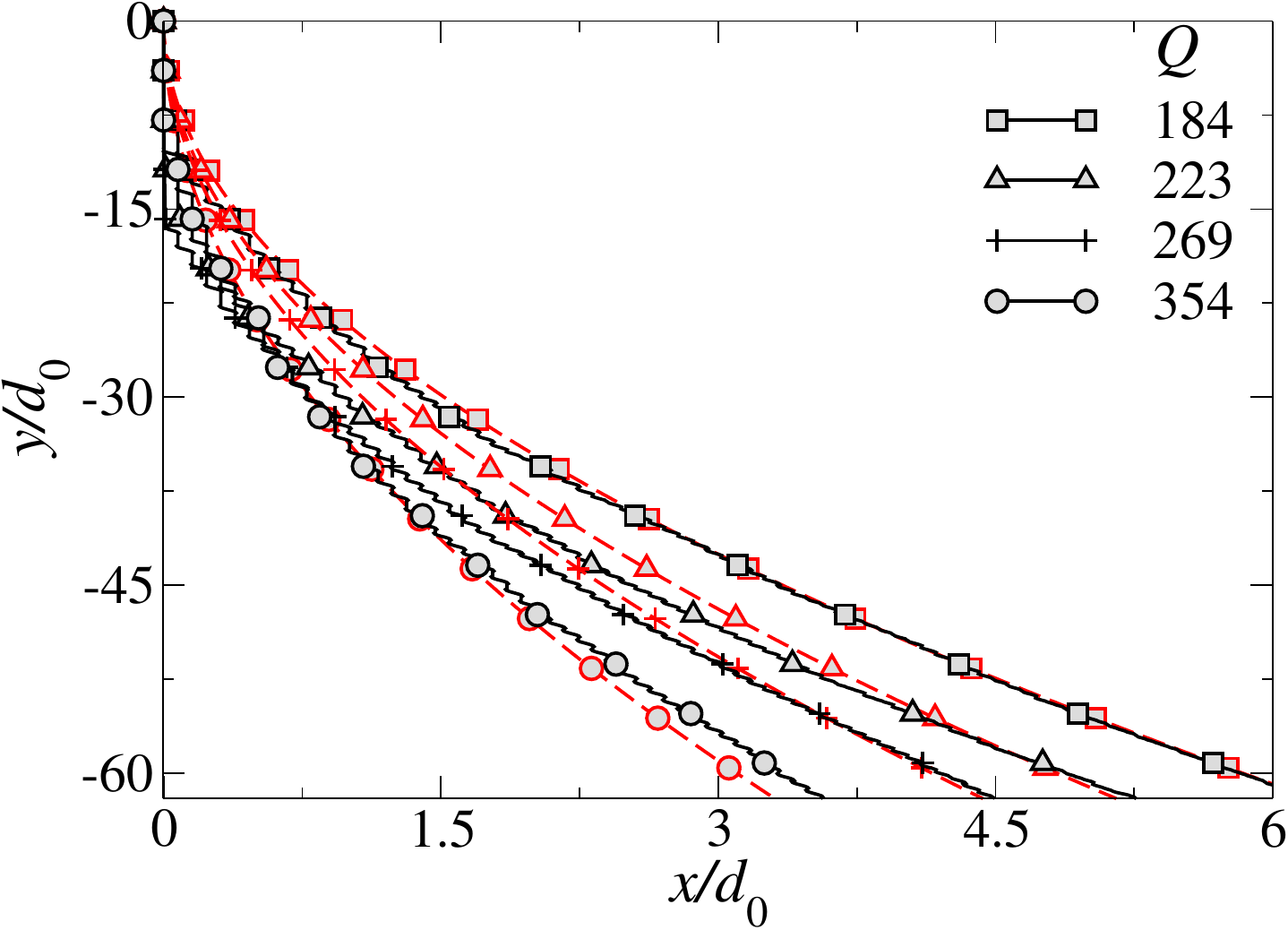}
\includegraphics[width=0.45\textwidth]{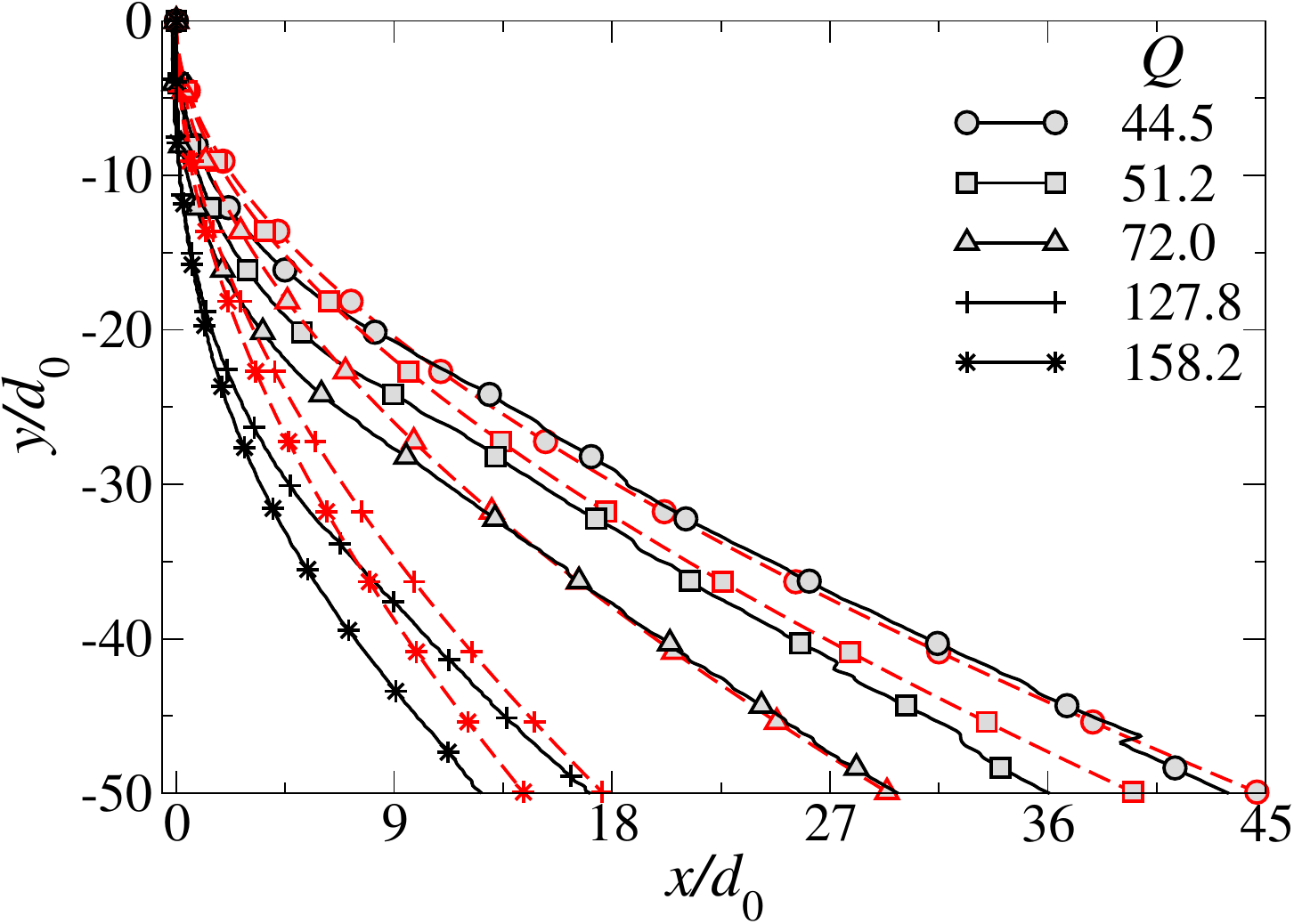}
\caption{Comparison of liquid jet trajectories predicted by the semi-empirical model (equation \ref{eqn:Kasmaiee}, red dashed lines) with experimental measurements (solid black lines) for different values of the momentum flux ratio $Q$ in the (a) column breakup regime and (b) bag breakup regime.}
\label{fig:traj_comparison}
\end{figure}

The trajectory of a liquid jet injected transversely into a crossflowing airstream can be described using a semi-empirical framework that captures the balance between liquid inertia and aerodynamic drag. In the present study, we adopt the model proposed by \cite{kasmaiee2022influence}. The formulation assumes that the drag is primarily governed by the streamwise velocity difference between the gas and liquid phases, while transverse velocity differences are neglected. Additionally, the jet is assumed to retain a constant diameter and a uniform transverse velocity component, with gravitational effects being negligible. The resulting jet trajectory is expressed as \citep{kasmaiee2022influence}
\begin{equation}
\frac{x}{d_0}
=
C_1
\left[
\frac{\sqrt{\rho_j/\rho_g}}{\sqrt{Q}\,\sin\theta}
\left(\frac{y}{d_0}\right)
-
\frac{1}{C_2}
\ln\!\left(
1
+
C_2
\left(
\frac{\sqrt{\rho_j/\rho_g}}{\sqrt{Q}}
-
\cot\theta
\right)
\left(\frac{y}{d_0}\right)
\right)
\right]
\mathrm{We}_g^{\,C_3},
\label{eqn:Kasmaiee}
\end{equation}
where $\theta$ denotes the jet injection angle (in the present study, $\theta=90^\circ$) and $Q \equiv We_j/We_g$ denotes the momentum flux ratio. The coefficients $C_1$, $C_2$, and $C_3$ are empirical constants introduced to account for deviations from the idealized assumptions of the model. In the present study, these constants are obtained via nonlinear least-squares fitting to the experimental data, yielding $C_1 = 0.61$, $C_2 = 0.006$, and $C_3 = 0.05$. Figure \ref{fig:traj_comparison} presents the comparison between the predicted trajectories (red dashed lines) and experimental measurements (solid black lines) for a range of $Q$, spanning both column (figure \ref{fig:traj_comparison}a) and bag breakup (figure \ref{fig:traj_comparison}b) regimes. In the column breakup regime (figure \ref{fig:traj_comparison}a), the model shows good agreement with the experimental data. As the value of $Q$ increases, both the predicted and measured trajectories exhibit increased penetration, reflecting the dominant influence of liquid inertia. Minor deviations observed in the downstream region can be attributed to the development of interfacial instabilities, which are not explicitly captured in the model. In the bag breakup regime (figure \ref{fig:traj_comparison}b), the model continues to predict the overall trend of the trajectories. However, noticeable discrepancies arise, particularly at lower values of $Q$. These deviations primarily result from the stronger influence of unsteady aerodynamic forces, large-scale deformations associated with bag formation, and the growth of Rayleigh–Taylor instabilities and surface undulations, effects that are not captured by the semi-empirical formulation.

Additionally, the breakup length of a liquid jet in a crossflow is a crucial parameter governing atomization and spray formation, as it defines the distance over which the liquid core remains intact before disintegrating into ligaments and droplets. In the present study, the breakup point is identified as the location where the continuous liquid column ceases to exist, and is characterized by the vertical breakup height $y_b$ and the streamwise breakup location $x_b$. The breakup process is dictated by the interplay between liquid inertia, aerodynamic forces, and surface tension. \cite{wu1997breakup} showed that the vertical breakup height follows $\frac{y_b}{d_0} \sim Q^{\frac{1}{2}}$. Consistent with this scaling, a best-fit correlation based on the present experimental data is given by
\begin{equation}
\frac{y_b}{d_0} = 4.67 ~ We_g^{-0.128} ~ Q^{0.5},
\label{eqn:breakup_length}
\end{equation}
with a high coefficient of determination ($R^2 = 0.97$). It can be observed that the exponent of $Q$ is in close agreement with the theoretical value of $1/2$, confirming that the breakup height is primarily governed by the balance between liquid inertia and aerodynamic forces. Since $Q=We_j/We_g$, the correlation can equivalently be expressed as
\begin{equation}
\frac{y_b}{d_0} = 4.67 ~ {We_j}^{0.5}  ~ {We_g}^{-0.628}.
\label{eqn:breakup_length}
\end{equation}
This relationship reveals that the vertical breakup location depends on both $We_g$ and $We_j$. The normalized streamwise breakup position is found to remain approximately constant, ${x_b/d_0} = 10.8 \pm 0.73$, indicating minimal variation with changes in $Q$ and $We_g$. This indicates that, unlike the vertical penetration, the streamwise breakup location is relatively insensitive to global flow conditions and is instead governed by local instability dynamics. This observation is consistent with the findings of \cite{wu1997breakup}, who reported that the horizontal breakup coordinate remains nearly invariant for normal injection into crossflow. Similar behaviour has also been reported in other experimental studies, where the streamwise breakup location shows significantly lower sensitivity to flow parameters compared to the vertical penetration \citep{sallam2004breakup, wang2012bag}.

\begin{figure}
\centering
\includegraphics[width=0.5\textwidth]{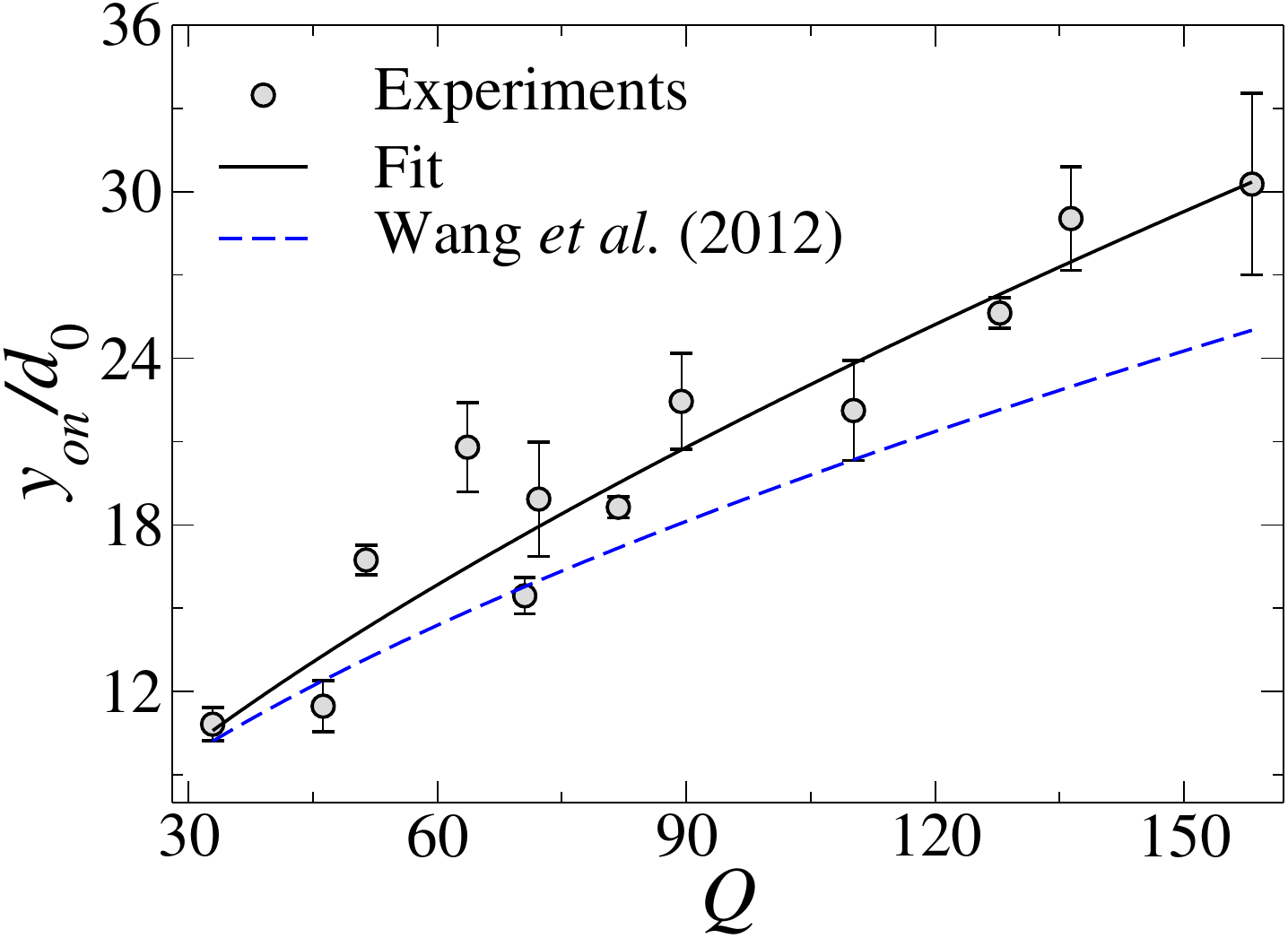}
\caption{Variation of the normalized bag onset height $\left(y_{on}/d_0\right)$ with the momentum flux ratio $Q$. The solid line represents the power-law fit given by eq.~(\ref{eqn:bag_onset}), while the symbols denote the present experimental data and the dashed line represents the results reported by \cite{wang2012bag}.}
\label{fig:bag_onset}
\end{figure}

Furthermore, the bag onset length ($y_{on}$) defines the location along the jet trajectory where the transition from a continuous liquid column to bag-type deformation first occurs. This point signifies the onset of large-scale interfacial instabilities, leading to the formation of inflated liquid membranes prior to breakup \citep{wang2012bag}. In the present study, the bag onset location is identified from shadowgraph images as the position where a distinct bag structure first emerges and is expressed in normalized form. The onset of bag formation is initiated by aerodynamic instabilities and is therefore governed primarily by the gas Weber number, $We_g$. As $We_g$ increases, the aerodynamic stresses acting on the jet surface intensify, promoting the growth of Rayleigh-Taylor instabilities and resulting in an earlier onset of bag formation \citep{kasmaiee2025backward}. In contrast, increasing the liquid Weber number ($We_j$) delays the onset because the enhanced liquid inertia resists jet deformation \citep{wu1997breakup}. Thus, although the instability is triggered primarily by the gas Weber number, the streamwise location at which the first bag appears is governed by the balance between aerodynamic forcing and liquid inertia. Accordingly, the bag-onset location is correlated using the momentum flux ratio, $Q= We_j/We_g$, and a least-squares fit to the present experimental data yields
\begin{equation}
\frac{y_{on}}{d_0} \approx 1.02~Q^{2/3}.
\label{eqn:bag_onset}
\end{equation}
As illustrated in figure~\ref{fig:bag_onset}, the proposed correlation provides a representation of the experimental data over the range of operating conditions investigated. It should be noted that the exponent of $2/3$ is obtained empirically from the present dataset and is not derived from a theoretical scaling analysis. Nevertheless, the findings of this section provide insight into the regions along the jet where the droplet size distribution undergoes pronounced variations, thereby enabling the selection of representative spatial locations for detailed characterization of the droplet size distributions.

\subsection{Drop size distribution}
\label{subsec:holography_dsd}

In this section, we examine the droplet size distributions produced by the fragmentation of a liquid jet in a crossflowing airstream under various governing conditions using digital in-line holography (DIH). A theoretical framework based on interfacial instability mechanisms together with a two-parameter Gamma distribution is employed to characterize the measured droplet size distributions. To the best of our knowledge, this is the first study to utilize such a framework to analyze droplets generated from a downward liquid jet in a crossflowing airstream at different spatial locations undergoing distinct fragmentation mechanisms, such as bag, rim, and node breakup. These mechanisms are governed by the competition between destabilizing aerodynamic forces and stabilizing surface tension. Bag breakup produces the finest droplets through the inflation and rupture of a thin liquid membrane, whereas rim and node breakup generate progressively larger droplets through the disintegration of thicker liquid structures at the jet periphery. The characteristic droplet sizes associated with each mechanism are derived in the following subsections and form the basis for constructing the overall droplet size distribution.

The droplet size distribution is quantified using the volume-based probability density function, $P_v$, which accounts for the volumetric contribution of droplets of a given size in a control volume \citep{villermaux2007fragmentation,jackiw2022prediction,ade2023size}. It is defined as
\begin{equation}
P_v = \frac{\chi^3 P_n(\chi)}{\int_{0}^{\infty} \chi^3 P_n(\chi) \d \chi}
= \frac{\chi^3 P_n(\chi)}{\beta^3 \Gamma(\alpha+3)/\Gamma(\alpha)},
\label{eqn:vpd}
\end{equation}
where $P_n(\chi)$ is the number-based probability density function given by
\begin{equation}
P_n(\chi) = \frac{\chi^{\alpha - 1} e^{-\chi/\beta}}{\beta^\alpha \Gamma(\alpha)}.
\label{npd}
\end{equation}
Here, $\chi = d/d_0$ is the normalized droplet diameter, $\Gamma$ denotes the Gamma function, and $\alpha$ and $\beta$ are the shape and rate parameters, respectively. These parameters are evaluated from the statistical moments of the distribution, which are given by \citep{jackiw2022prediction}
\begin{equation}
\alpha = \left(\frac{\bar{\chi}}{s}\right)^2, \quad
\beta = \frac{s^2}{\bar{\chi}},
\label{alpha_beta}
\end{equation}
where $\bar{\chi}$ and $s$ represent the mean and standard deviation of the normalized droplet size, respectively. It should be emphasized that the Gamma distribution is purely empirical, where the parameters $\alpha$ and $\beta$ represent the skewness and the spread of the distribution, respectively. The characteristic droplet sizes associated with the different breakup mechanisms are estimated, thereby establishing a direct connection between the underlying fragmentation processes and the resulting droplet size distributions.

\subsubsection{Bag breakup} \label{subsubsec:bag}

Bag rupture produces droplets spanning a range of characteristic sizes. The smallest droplets originate from the rupture of the thin liquid membrane and are governed by the Taylor–Culick velocity \citep{culick1960comments}:
\begin{equation}
d_B = \frac{2\sigma}{\rho_j u_{rr}^2},
\label{d_B}
\end{equation}
where $u_{rr}$ is the receding rim velocity. The thickness of the receding rim itself defines a second characteristic length scale, which is given by
\begin{equation}
d_{rr,B} = \sqrt{\frac{\sigma}{\rho_j a_c}},
\label{d_rrB}
\end{equation}
where $a_c = u_{rr}^2/R_f$ is the deceleration of the receding rim, wherein $R_f$ is the radius of the bag at the onset of burst. This rim is disintegrated due to Rayleigh–Plateau instability, producing droplets of size
\begin{equation}
d_{RP,B} = 1.89~ d_{rr,B}.
\label{d_RPB}
\end{equation}
Subsequent ligament breakup yields smaller satellite droplets, given by
\begin{equation}
    d_{sat,B} = \frac{d_{RP,B}}{\sqrt{2 + 3~Oh_{rr}/\sqrt{2}}},
    \label{d_satB}
\end{equation}
where $Oh_{rr}$ is the Ohnesorge number based on the rim thickness.

\subsubsection{Rim breakup} \label{subsubsec:rim}

Following bag rupture, the main rim disintegrates through Rayleigh–Plateau instability and receding rim dynamics. The primary droplet size is set by the final rim thickness $h_f$,
\begin{equation}
d_R = 1.89~h_f,
\label{d_R}
\end{equation}
while receding rim instability produces droplets of size
\begin{equation}
d_{rr} = d_0 \left[ 1.5 \left(\frac{h_f}{d_0}\right)^{2} \frac{\lambda_{rr}}{d_0} \right]^{1/3},
\label{d_rr}
\end{equation}
wherein $\lambda_{rr} = 4.5\,d_{rr,B}$ denotes the wavelength of the receding rim instability. The viscous effects during ligament breakup further modify these characteristic sizes, leading to the formation of satellite droplets associated with the main and receding rims, respectively, given by
\begin{equation}
d_{sat,R} = \frac{d_R}{\sqrt{2 + 3 Oh_R/\sqrt{2}}}, \qquad
d_{sat,rr} = \frac{d_{rr}}{\sqrt{2 + 3 Oh_R/\sqrt{2}}},
\label{d_sat_R}
\end{equation}
where $Oh_R$ is the Ohnesorge number based on $h_f$.

\subsubsection{Node breakup} \label{subsubsec:node}

Nodes formed at the rim periphery undergo breakup under combined Rayleigh–Taylor and Rayleigh–Plateau instabilities \citep{zhao2010morphological, ade2023size}. The characteristic droplet size is given by \citep{jackiw2022prediction}
\begin{equation}
d_N = d_0\left[1.5\left(\frac{h_i}{d_0}\right)^{2}\frac{\lambda_{RT}}{d_0} n\right]^{1/3},
\label{eq:d_N}
\end{equation}
where $\lambda_{RT}$ denotes the maximum wavelength of the RT instability and $n$ is the volume fraction of the nodes relative to the liquid disk. 

Further details of the theoretical analysis can be found in \cite{ade2023size,jackiw2022prediction}.

\subsubsection{Surface nodes} \label{subsubsec:surface_nodes}

In addition to the nodes shed from the periphery of the main rim during bag and rim rupture, a distinct population of nodes develops on the windward surface of the liquid column due to RT instability. Unlike the rim-peripheral nodes, whose characteristic size is governed by the RT wavelength ($\lambda_{RT}$) associated with the deforming rim, these surface nodes arise from the amplification of small-amplitude interfacial disturbances generated by the interaction between the liquid jet and the high-speed crossflow. Their characteristic wavelength is therefore controlled by the gas Weber number and the effective thickness of the liquid surface layer subjected to aerodynamic forcing. For a system involving two fluids of different densities $(\rho_j, \rho_g)$, the growth rate $(\eta)$ of a small interfacial disturbance with wavenumber $(k)$ is given by \citep{chandrasekhar2013hydrodynamic}:
\begin{equation}
 \eta^2 = ak\left[\frac{\rho_j - \rho_g}{\rho_j + \rho_g} - \frac{k^2\sigma}{a(\rho_j + \rho_g)}\right],
 \label{eq:chandrasekhar}
 \end{equation}
where $a$ is the interfacial acceleration. Since $\rho_j \gg \rho_g$ for the present liquid-gas system, eq.~\eqref{eq:chandrasekhar} simplifies to
\begin{equation}
 \eta^2 \approx ak - \frac{k^3\sigma}{\rho_j}.
 \label{eq:growth_simplified}
 \end{equation}
Unlike column-wave models \citep{ng2008bag}, where the entire liquid column responds to aerodynamic forcing, the present formulation assumes that only a thin liquid layer near the interface participates in the instability dynamics \citep{sinha2019surface}. Denoting the effective thickness of this responsive layer by $h$, the aerodynamic pressure balance at the interface can be expressed as
\begin{equation}
\frac{1}{2}\rho_g v_g^2 = \rho_j a h.
\label{eq:pressure_balance}
\end{equation}
Substituting eq.~\eqref{eq:pressure_balance} into eq.~\eqref{eq:growth_simplified} yields
\begin{equation}
 \eta = \left(\frac{k\rho_g v_g^2}{2\rho_j h}
 - \frac{k^3\sigma}{\rho_j}\right)^{1/2}.
 \label{eq:growth_full}
 \end{equation}

The most amplified wavenumber is obtained by maximizing eq.~\eqref{eq:growth_full} with respect to $k$, which is given by
\begin{equation}
k = \sqrt{\frac{\rho_g v_g^2}{6\sigma h}}.
\label{eq:kmax}
\end{equation} 
Using $k = 2\pi/\lambda$, the dominant surface-wave wavelength becomes
\begin{equation}
\lambda = 2\pi\sqrt{\frac{6\sigma h}{\rho_g v_g^2}},
\label{eq:lambda}
\end{equation}
which can be rewritten in terms of the gas Weber number $\We_g$ as \citep{sinha2019surface}
\begin{equation}
\frac{\lambda}{d_0} = 2\pi\sqrt{\frac{6}{\We_g}\frac{h}{d_0}}.
\label{eq:lambda_nd}
\end{equation}
The instability induced by the crossflow is primarily confined to a thin liquid layer near the gas–liquid interface rather than involving the entire liquid column. Consequently, only a finite surface region actively participates in the early instability dynamics. Following \cite{wang2014theoretical}, the effective responsive layer thickness is approximated as $h=d_0/16$, which provides a reasonable estimate of the dominant instability wavelength and the resulting breakup structures in jet-in-crossflow atomization. Thus, eq.~\eqref{eq:lambda_nd} becomes
\begin{equation}
\frac{\lambda}{d_0} = 2\pi\sqrt{\frac{3}{8\We_g}}.
\label{eq:lambda_nd_compiled}
\end{equation}
Equation~\eqref{eq:lambda_nd_compiled} shows that the characteristic surface-wave wavelength decreases with increasing $\We_g$, reflecting the stronger aerodynamic forcing at higher crossflow velocities. For example, based on the present experimental observations, $\lambda/d_0 = 1.36$ for $\We_g = 8$, whereas $\lambda/d_0 = 0.84$ for $\We_g = 21$. Moreover, the characteristic size of the surface-generated nodes differs from that of the bag rim-peripheral nodes governed by $\lambda_{RT}$. This distinction must be incorporated into the overall droplet size distribution, together with the combined contributions from bag, rim, and node breakup processes, as discussed in the following section.

\begin{figure}
\centering
\includegraphics[width=0.95\textwidth]{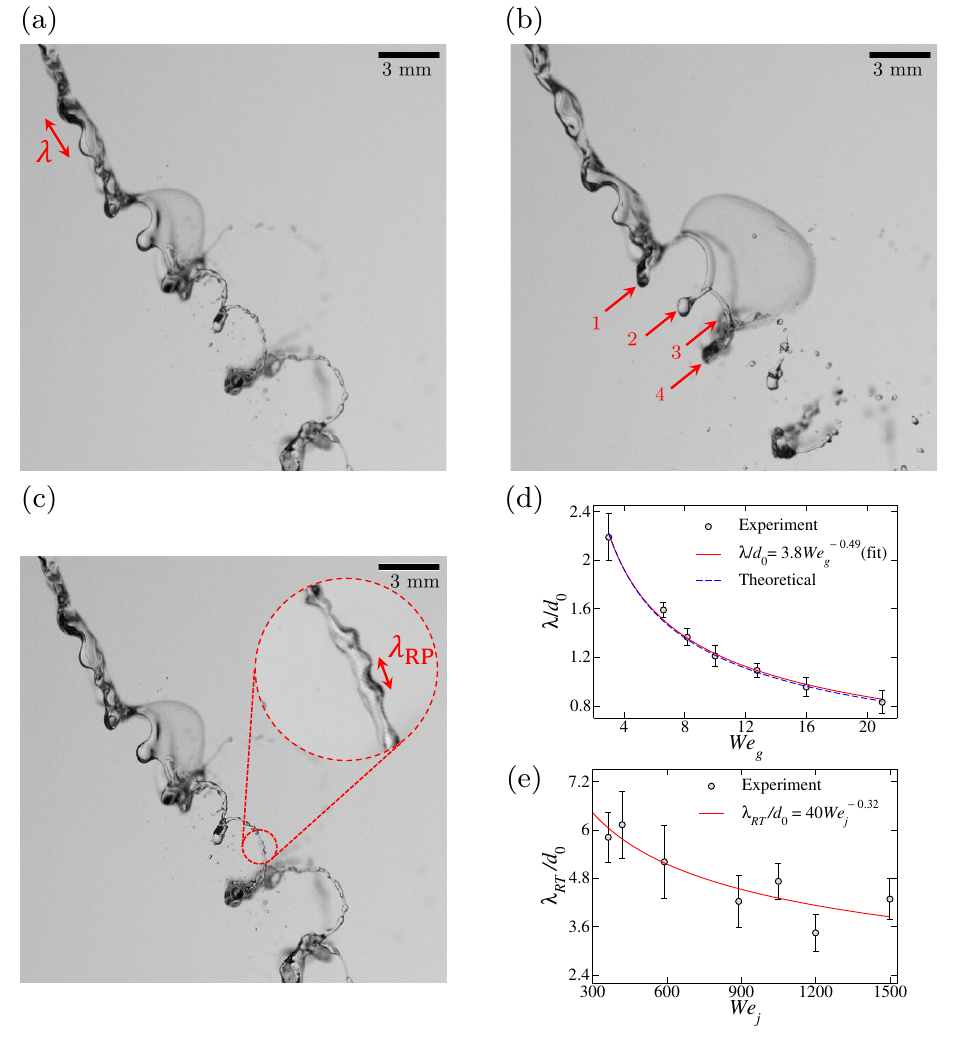}
\caption{Liquid jet in a uniform crossflow at $We_j \approx 420$ and $We_g \approx 8$ illustrating the characteristic instability wavelengths: (a) Rayleigh-Taylor (RT) wavelength of the surface waves on the liquid column; (b) RT nodes formed along the bag rim; (c) Rayleigh-Plateau (RP) wavelength associated with the breakup of the bag rim. (d) Variation of the  wavelength of the column disturbances with $We_g$ at a constant jet Weber number, $We_j \approx 420$. (e) Variation of the RT wavelength of the bag-rim nodes with $We_j$ at a constant gas Weber number, $We_g \approx 8$.}
\label{fig:wavelegth_RT_RP_Nodes}
\end{figure}

To further elucidate the instability mechanisms underlying the observed breakup modes, we quantify the characteristic wavelengths of the interfacial deformations. Figure~\ref{fig:wavelegth_RT_RP_Nodes}(a-c) presents shadowgraphy images depicting the different instability modes. Specifically, the wavelengths of the surface waves on the jet column ($\lambda$), the Rayleigh-Taylor (RT) instability associated with the nodes formed on the bag rim ($\lambda_{RT}$), and the Rayleigh-Plateau (RP) instability of the bag rim ($\lambda_{RP}$) are also shown in figure~\ref{fig:wavelegth_RT_RP_Nodes}(a-c), respectively. The RT wavelength, $\lambda_{RT}$, is estimated indirectly from the number of nodes formed along the bag rim following \citet{zhao2010morphological}. The relationship is given by $N_{RT}={2\pi R_{\max}/\lambda_{RT}}$, where $N_{RT}$ denotes the number of nodes on the bag rim, as illustrated in figure \ref{fig:wavelegth_RT_RP_Nodes}(b), and $R_{\max}$ is the maximum radius of the bag rim. The variation of  $\lambda/d_{0}$ with $We_g$, and the variation of the measured $\lambda_{RT}/d_{0}$ with $We_j$ is presented in figure~\ref{fig:wavelegth_RT_RP_Nodes}(d) and \ref{fig:wavelegth_RT_RP_Nodes}(e), respectively. As shown in figure~\ref {fig:wavelegth_RT_RP_Nodes}(a) and \ref {fig:wavelegth_RT_RP_Nodes}(d), the measured wavelength of the disturbances($\lambda$) developing on the jet column is in close agreement with the wavelength of the fastest-growing RT mode obtained from eq.~\eqref{eq:lambda_nd_compiled}, indicating that the initial deformation of the liquid column is governed by RT instability induced by the aerodynamic forcing of the crossflow. Furthermore, the disturbance wavelength decreases with increasing $We_g$ and follows a power-law relationship, as shown in figure~\ref{fig:wavelegth_RT_RP_Nodes}(d). Conversely, we observe that $We_j$ has a negligible influence on the wavelength of the column disturbances (not shown). As these RT waves are convected downstream, their amplitudes continue to grow, ultimately leading to the disintegration of the liquid column. As discussed in the earlier sections, the aerodynamic loading progressively flattens and stretches the column, inflating it into a bag-like structure. The initiated bag subsequently develops node-like structures along the rim due to RT instability. In contrast to the fragmentation of an isolated droplet, where the RT wavelength is primarily determined by the gas Weber number, the bag breakup of a jet in crossflow is also influenced by the liquid inertia through the jet Weber number, $We_j$, as shown in figure~\ref{fig:wavelegth_RT_RP_Nodes}(e). Consequently, the measured RT wavelength decreases as $We_j$ increases, reflecting enhanced stretching and thinning of the liquid column at higher liquid inertia. The theoretical estimate for $\lambda_{RT}$ is given by \citep{jackiw2022prediction}
\begin{equation}
\frac{\lambda_{RT}}{d_0} = \frac{4\pi}{\sqrt{C_DWe_{g}}}\left(\frac{d_0}{2R_{max}}\right),
\end{equation}
where $C_{D}=1.2$ is the drag coefficient of the disk-like structure formed due to aerodynamic forcing. We found that the theoretical prediction yields $\lambda_{RT}/d_0 \approx 4.46$, whereas the experimentally measured values lie in the range of $4.9$-$6.3$ for $We_j=420$, demonstrating reasonable agreement between the model and the measurements. Similarly, the characteristic wavelength measured along the bag rim agrees well with the dominant RP wavelength predicted by the theoretical modeling. The theoretical value of $\lambda_{RP}$ is given by \citep{jackiw2021aerodynamic,zhao2011breakup, obenauf2021theoretical}
\begin{equation}
\frac{\lambda_{RP}}{d_0} = 4.5\frac{h_f}{d_{0}},
\end{equation}
where $h_f$ is the final rim thickness (see details in \cite{jackiw2021aerodynamic}). The theoretical prediction gives $\lambda_{RP}/d_0 \approx 0.61$, while the experimentally measured value is approximately $0.55$ for $We_j=420$ and $We_g=8$. The close agreement confirms that the breakup of the bag rim is governed predominantly by the RP instability. The subsequent amplification of these capillary disturbances results in ligament formation, followed by the generation of secondary droplets, consistent with the classical RP breakup mechanism of bag-rim. It should be noted that the RT wavelength associated with the rupture of the thin liquid bag cannot be measured directly due to the highly transient nature of the rupture and the limited spatial resolution of the imaging system. Overall, the wavelength measurements presented in figure \ref{fig:wavelegth_RT_RP_Nodes}(a-e), together with the observed breakup morphologies, provide quantitative evidence supporting the proposed sequence of instability mechanisms governing the breakup of a liquid jet in crossflow.

\subsection{Droplet size distribution along the jet trajectory}
\label{subsec:jet_breakup_dsd}

The spatial evolution of the droplet size distribution provides physical insight into the sequence of instability mechanisms governing the fragmentation of a liquid jet in a crossflowing airstream. Since bag, rim, and node breakup processes produce droplets with distinct characteristic length scales (as discussed in \S\ref{subsec:holography_dsd}), the measured distributions along the jet trajectory reflect the local dominance of each breakup mode and their interactions. This establishes a direct relationship between the underlying instability mechanisms and the observed statistical characteristics of the droplet size distribution. The droplet size distributions are obtained using digital in-line holography (DIH) at discrete locations along the jet trajectory. The holographic measurements are performed over a field of view of approximately $20.5~\mathrm{mm} \times 20.5~\mathrm{mm}$ (corresponding to the camera resolution reported in \S\ref{subsec:DIH}), reconstructed over a depth of $140~\mathrm{mm}$ in the $z$-direction, and averaged over approximately 100 realizations. The two-pixel detection threshold sets a minimum detectable droplet diameter of $\sim 40~\mu\text{m}$. Consequently, droplets smaller than this size are excluded from the reported distributions. The omission of droplets below $40~\mu\text{m}$ primarily affects the low-diameter tail of the probability density function (PDF). However, the characteristic droplet sizes produced by the dominant breakup modes investigated in this study except the length scales corresponding to the bag thickness lie well above this detection limit. Therefore, the peak locations, overall shape, and relative contributions of the different breakup mechanisms to the PDFs are not significantly affected by the exclusion of these smaller droplets. All spatial locations are referenced to the needle exit, defined as the origin in the coordinate system, as shown in figure \ref{fig:shadowgraphy_schematic}. These locations are selected along the curved jet trajectory to capture the progressive transition of the jet from primary deformation to fully developed atomization. Furthermore, the theoretical analysis outlined in \S\ref{sec:theory} facilitates the identification of regions along the jet where the droplet size distribution exhibits significant variation, thereby guiding the selection of appropriate locations for detailed holographic measurements. 

\begin{figure}
\centering
\includegraphics[width=0.85\textwidth]{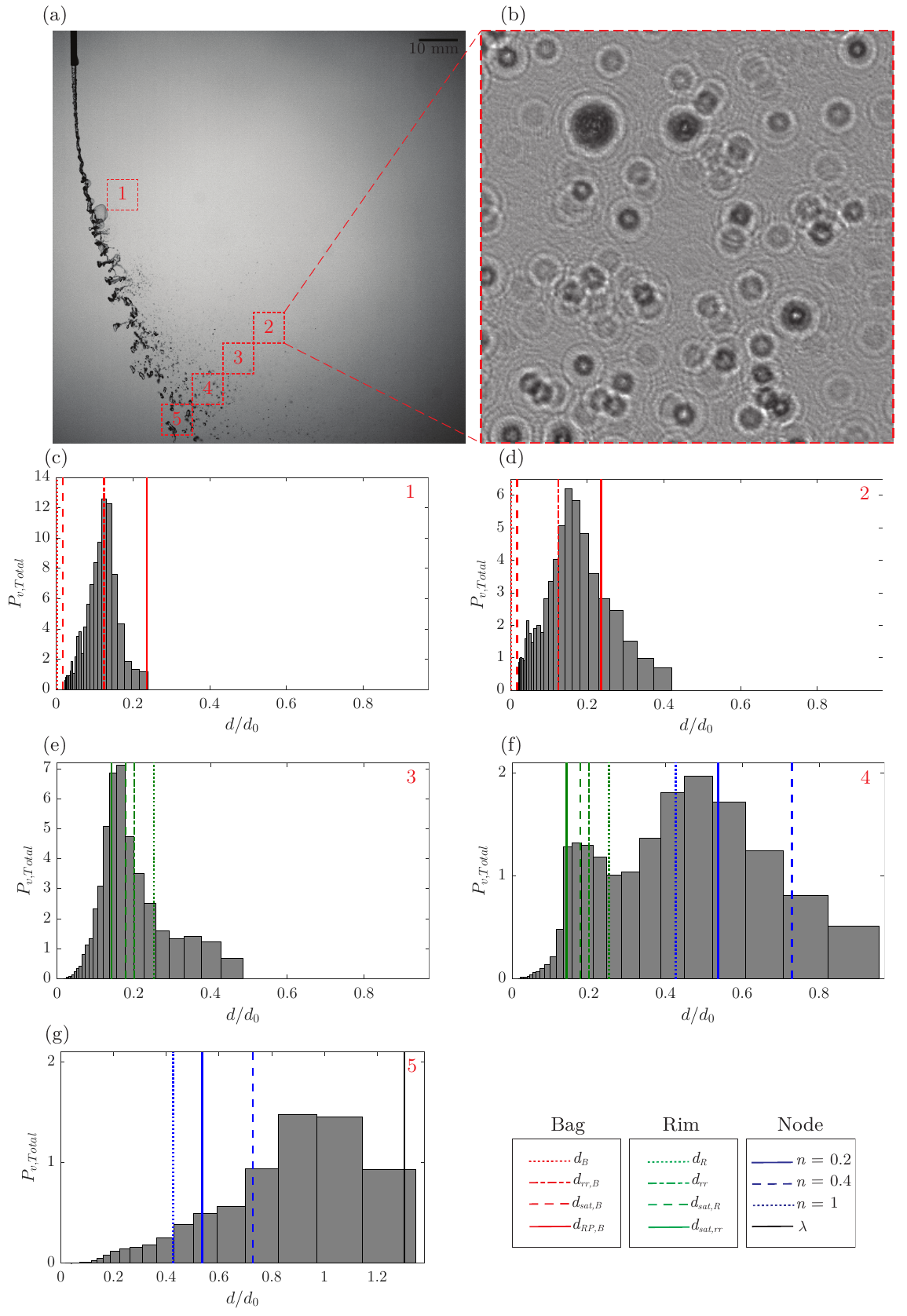}
\caption{Comparison of the theoretical characteristic droplet sizes with the DIH measurements for $We_j = 1511$ and $We_g = 8$. (a) Shadowgraph image of a liquid jet undergoing fragmentation, with regions 1–5 selected for DIH-based measurements. The holographic measurements are performed over a field of view of approximately $20.5~\mathrm{mm} \times 20.5~\mathrm{mm}$ (see \S2.3), with the measurement locations referenced to the needle exit at $(0,0)$. The centers of regions 1–5 are located at $(x/d_0, y/d_0) = (9.21, 31)$, $(37.5, 59)$, $(30.9, 65.6)$, $(24.41, 72.2)$, and $(17.85, 78.7)$, respectively. (b) Representative hologram corresponding to region 2 in panel (a). Panels (c)–(g) present the volume-based probability density functions, $P_{v,\mathrm{Total}}$, as a function of the normalized droplet diameter, $d/d_0$, for regions 1–5. The vertical lines of different colors in each panel denote the theoretical characteristic droplet sizes associated with bag, rim, and node breakup mechanisms, as indicated in the legends at the bottom-right corner of the figure.}
\label{fig:dsd_We8}
\end{figure}

Figure \ref{fig:dsd_We8} illustrates the spatial evolution of the droplet size distribution along the trajectory of a liquid jet undergoing fragmentation at $We_g = 8$ and $We_j = 1511$. Figure \ref{fig:dsd_We8}(a) presents a shadowgraph image of the jet, where regions 1–5 are identified as representative locations for droplet size measurements. These regions are centered at $(x/d_0, y/d_0) = (9.2, 31)$, $(37.5, 59)$, $(30.9, 65.6)$, $(24.4, 72.2)$, and $(17.8, 78.7)$, respectively. These selected regions for DIH represent successive stages of the fragmentation process, corresponding approximately to bag-rupture-dominated breakup (region 1), combined bag and rim fragmentation (region 2), rim-dominated breakup (region 3), coexistence of rim and node disintegration (region 4), and node/surface-wave-driven fragmentation (region 5). A representative hologram corresponding to region 2 is shown in figure \ref{fig:dsd_We8}(b), demonstrating the quality of the measurements and the droplet field captured using DIH. The corresponding droplet size distributions for the different regions are presented in figure \ref{fig:dsd_We8}(c–g), where the volume-based probability density, $P_{v,\mathrm{Total}}$, is plotted as a function of the normalized droplet diameter, $d/d_0$. The distributions exhibit a systematic spatial evolution of droplet sizes along the jet trajectory, reflecting the progression and interaction of the underlying breakup mechanisms. The theoretical characteristic droplet sizes associated with the different breakup processes are indicated by distinct lines in the bottom-right panel of figure \ref{fig:dsd_We8}.

In the near-field (region 1, figure \ref{fig:dsd_We8}c), the droplet size distribution is dominated by droplets generated through continuous bag rupture. The peak at $d/d_0 \approx 0.12$ corresponds to droplets formed from the rupture of the thin liquid membrane and the subsequent Rayleigh–Plateau instability of the receding rim. The relatively narrow distribution reflects the well-defined thickness of the inflated liquid sheet, which sets the dominant length scale in this region. Accordingly, both the mean droplet size and the standard deviation remain small. The corresponding Gamma distribution is characterized by a relatively large $\alpha ~(\approx 9.4)$ and a small $\beta ~(\approx 0.012)$, indicating a narrow distribution. Closer examination further shows that the theoretically predicted characteristic droplet size associated with bag breakup (indicated by the vertical lines in figure \ref{fig:dsd_We8}) lies well within the experimentally observed distribution, confirming that bag breakup is the dominant fragmentation mechanism in this region. Further downstream, in region 2, the distribution broadens and shifts slightly toward larger diameters, indicating the accumulation and downstream convection of droplets generated from successive bag breakup and rim fragmentation events. This evolution is accompanied by a reduction in $\alpha ~(\approx 4.7)$ and an increase in $\beta ~(\approx 0.04)$, consistent with a broader distribution and an increase in the characteristic droplet size. In this region, the experimentally observed characteristic droplet size exceeds the theoretical prediction based solely on pure bag breakup, indicating the combined influence of both bag and rim breakup mechanisms in the measured distributions.

Further downstream, in region 3 (figure \ref{fig:dsd_We8}e), the distribution shifts toward larger droplet sizes, consistent with the onset of rim-dominated breakup. In this regime, the characteristic droplet size is governed by the thickness of the liquid rim and its destabilization through the Rayleigh–Plateau instability. The dominant droplet sizes agree well with the theoretical predictions for rim breakup. However, the distribution becomes increasingly skewed and asymmetric, reflecting the coexistence of droplets originating from residual bag fragments and newly formed ligaments. Correspondingly, $\beta ~(\approx 0.048)$ increases further, while $\alpha ~(\approx 4.3)$ decreases, indicating large variability in the droplet sizes. In region 4 (figure \ref{fig:dsd_We8}f), the distribution exhibits a bimodal character, indicating the simultaneous presence of rim and node breakup mechanisms. The larger-size peak corresponds to droplets generated through node disintegration at the rim periphery, whereas the smaller-size peak is associated with ligament and rim breakup. This behavior is also consistent with the theoretical predictions of the dominant characteristic droplet sizes associated with the respective fragmentation mechanisms.

\begin{figure}
\centering
\includegraphics[width=0.85\textwidth]{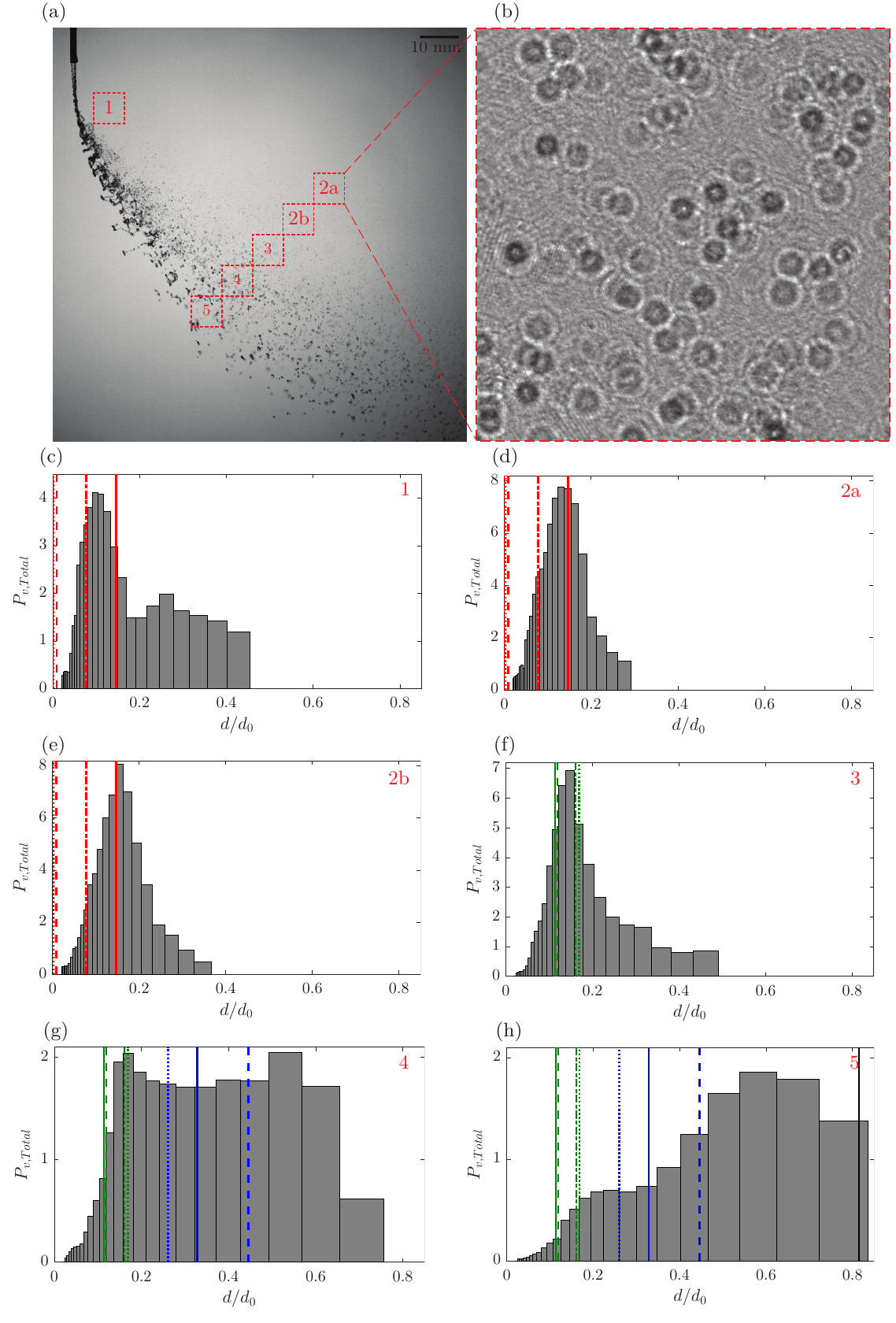}
\caption{Comparison of the theoretical characteristic droplet sizes with the DIH measurements for $We_j = 1511$ and $We_g = 21$. (a) Shadowgraph image of a liquid jet undergoing fragmentation, with different regions selected for DIH-based measurements. The measurements are conducted over a field of view of approximately $20.5~\mathrm{mm} \times 20.5~\mathrm{mm}$ (see \S2.3), with spatial locations referenced to the needle exit at $(0,0)$. The centers of regions 1, 2a, 2b, and 3–5 are located at $(x/d_0, y/d_0) = (6.2, 10.2)$, $(53, 27.9)$, $(46.6, 34.4)$, $(40, 40.9)$, $(33.5, 47.5)$, and $(26.9, 54)$, respectively. (b) Representative hologram corresponding to region 2a in panel (a). Panels (c)–(h) present the corresponding volume-based probability density functions, $P_{v,\mathrm{Total}}$, as a function of $d/d_0$. The vertical lines of different colors in each panel denote the theoretical characteristic droplet sizes associated with bag, rim, and node breakup mechanisms, as indicated in the legends at the bottom-right corner of figure \ref{fig:dsd_We8}.}
\label{fig:dsd_We21}
\end{figure}

In the far field (region 5), the distribution is dominated by the largest droplets, which originate from the breakup of node-like structures and surface corrugations formed on the windward side of the jet. These structures are associated with the Rayleigh–Taylor instability, whose characteristic wavelength $(\lambda)$ determines the dominant droplet size. The nodes and ligaments further undergo secondary fragmentation, producing intermediate-sized droplets with characteristic sizes smaller than $\lambda$. Consequently, the distribution becomes broader and more polydisperse due to the coexistence of multiple fragmentation pathways. The increasing polydispersity downstream arises from the simultaneous coexistence of multiple fragmentation mechanisms producing droplets over a broad range of characteristic length scales. Therefore, the resulting droplet population reflects the superposition of instability-driven breakup processes occurring over different spatial and temporal scales. In this region, both $\alpha ~(\approx 10.6)$ and $\beta ~(\approx 0.085)$ attain higher values, reflecting the increase in the mean droplet size together with the broadening of the distribution arising from the simultaneous action of multiple breakup mechanisms. The spatial variation in droplet populations across regions 2–5 represents the final stage of the fragmentation cascade, in which large-scale liquid structures fragment into discrete droplets through various instability-driven processes, accompanied by convective transport.

The corresponding number-based probability density ($P_n$) for the five DIH measurement regions shown in figure \ref{fig:dsd_We8}(a) are presented in figure \ref{fig:npd_We8}(a-h) in Appendix \ref{sec:A1}. Unlike the volume-based probability density function, $P_{v,\mathrm{Total}}$, which is weighted towards larger droplets, $P_n$ emphasizes the population of smaller droplets. In all regions, $P_n$ exhibits a pronounced peak at small normalized droplet diameters, indicating that the spray is dominated by fine droplets. Region 1 shows the highest population of small droplets due to active bag breakup. As the spray evolves downstream from regions 2 to 5, the distributions broaden and the population of fine droplets gradually decreases, reflecting the continued fragmentation of the liquid jet and ligaments. Nevertheless, fine droplets remain the dominant component of the spray in terms of number. Comparison of figures \ref{fig:dsd_We8} and \ref{fig:npd_We8} demonstrates the complementary nature of the two distributions. Here, $P_{v,\mathrm{Total}}$ is dominated by the liquid-volume contribution of larger droplets, and $P_n$ reveals that the overwhelming majority of droplets are much smaller.

Figure \ref{fig:dsd_We21} illustrates the spatial evolution of droplet size distributions along the jet trajectory at the higher gas Weber number, $We_g = 21$, corresponding to a regime of stronger aerodynamic forcing and more intense atomization. To isolate the effect of aerodynamic forcing, the jet Weber number is kept the same as in figure \ref{fig:dsd_We8} ($We_j = 1511$). The shadowgraph in panel (a) identifies multiple measurement regions, including additional intermediate zones (regions 2a and 2b), highlighting the increased spatial extent and complexity of the breakup process compared to the lower $We_g$ scenario. Figure \ref{fig:dsd_We21}(b) shows a representative hologram from an intermediate region, revealing a dense and highly dispersed droplet field indicative of rapid and vigorous fragmentation. The droplet size distributions presented in figure \ref{fig:dsd_We21}(c–h), expressed in terms of the volume-based probability density $(P_{v,\mathrm{Total}})$ as a function of the normalized diameter $(d/d_0)$, exhibit broader profiles across all regions compared to those at $We_g = 8$ (figure \ref{fig:dsd_We8}). The most prominent feature is that the droplet size distributions are systematically broader and shifted toward smaller droplet sizes across all regions compared to figure \ref{fig:dsd_We8}. This indicates enhanced atomization, where stronger aerodynamic forces promote earlier and more intense breakup. In the near-field (figure \ref{fig:dsd_We21}c), the distribution is significantly broader with a reduced peak magnitude, corresponding to lower $\alpha ~(\approx 3.0)$ and higher $\beta ~(\approx 0.07)$ values. This region is deminated by bag fragmentation close to the nozzle. Figures \ref{fig:dsd_We21}(d,e) illustrate the droplet size distributions in regions 2a and 2b, arising from the downstream convection of droplets generated through bag and rim fragmentation. The relatively narrow distributions observed in these regions suggest that the primary breakup mechanisms, namely bag rupture and rim breakup, occur simultaneously at much earlier stages and with greater intensity than in the lower-$We_g$ case.

As the flow progresses downstream, the distributions remain broad but continue to be biased toward smaller droplet sizes relative to the lower $We_g$ case, indicating reduced characteristic length scales (figure \ref{fig:dsd_We21}f). The appearance of multiple peaks or flattened profiles in certain regions (e.g., figure \ref{fig:dsd_We21}g) points to the coexistence of multiple breakup modes, including rim, node and ligament-mediated fragmentation. The persistence of fine droplets across downstream locations also indicates strong convective transport coupled with sustained secondary breakup. In contrast to figure \ref{fig:dsd_We8}, which exhibits a relatively orderly downstream transition from fine to coarse droplets, figure \ref{fig:dsd_We21}(h) demonstrates a more intense, spatially distributed, and fine-scale atomization process. Overall, the results highlight the critical role of aerodynamic forcing in governing the multiscale breakup dynamics and the resulting droplet statistics, with higher $We_g$ leading to earlier breakup, increased polydispersity, and a systematic shift toward finer droplets. Figure \ref{fig:npd_We21}(a-h) in the Appendix \ref{sec:A1} presents the corresponding number-based probability density functions ($P_n$) for the measurement regions shown in figure \ref{fig:dsd_We21}(a). Additionally, to examine the sensitivity of $P_n$ to the spatial resolution limit of $40~\mu$m, we perform a high-resolution DIH measurement with a resolution of $20~\mu$m for the bag-breakup event corresponding to Region 2(a) in figure~\ref{fig:dsd_We21}. The results (not shown) indicate that the resolution limit primarily affects the lowest-diameter bin, while the overall distribution remains unchanged for larger droplets. Furthermore, among the characteristic length scales associated with the breakup mechanisms considered, only the bag-thickness scale is below $40~\mu$m. All other characteristic scales exceed this limit and are therefore resolved in the present study. Similar to figure \ref{fig:npd_We8}(a-g), the distributions exhibit pronounced peaks at small normalized droplet diameters, indicating that the spray is predominantly dominated by small droplets. Owing to the higher gas Weber number, the distributions are shifted towards smaller droplet diameters, reflecting the enhanced atomization and increased production of fine droplets. As the spray evolves downstream, the distributions broaden slightly, while fine droplets remain the dominant component of the spray in terms of number. A comparison of figures \ref{fig:dsd_We21} and \ref{fig:npd_We21} again highlights the complementary nature of the volume- and number-based distributions. These findings provide a comprehensive characterization of the spatial evolution of the droplet size distribution during atomization.

\begin{figure}
\centering
\includegraphics[width=0.95\textwidth]{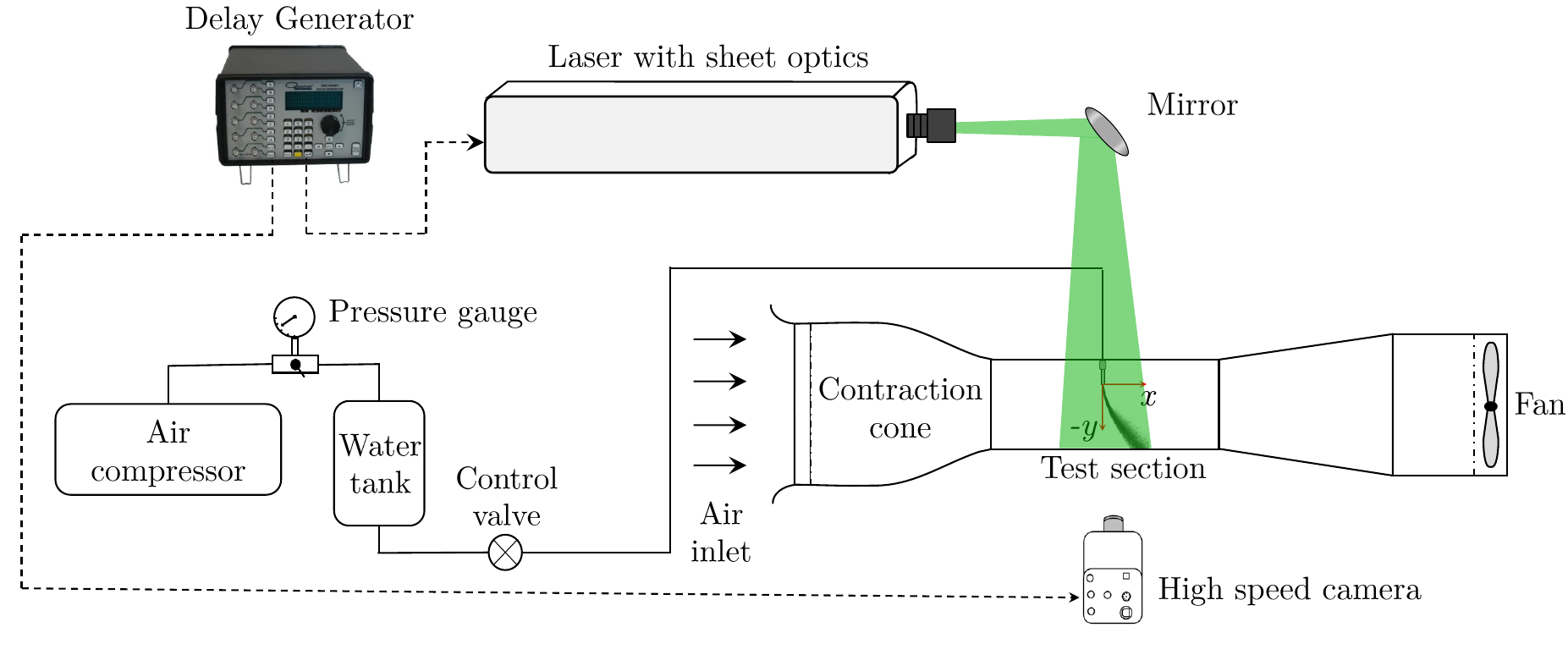}
\caption{Schematic of the PLIF (Planar Laser-Induced Fluorescence) setup used for droplet tracking during the breakup of a liquid jet in a crossflowing airstream.}
\label{fig:PLIF_schematic}
\end{figure}

\begin{figure}
\centering
\includegraphics[width=0.95\textwidth]{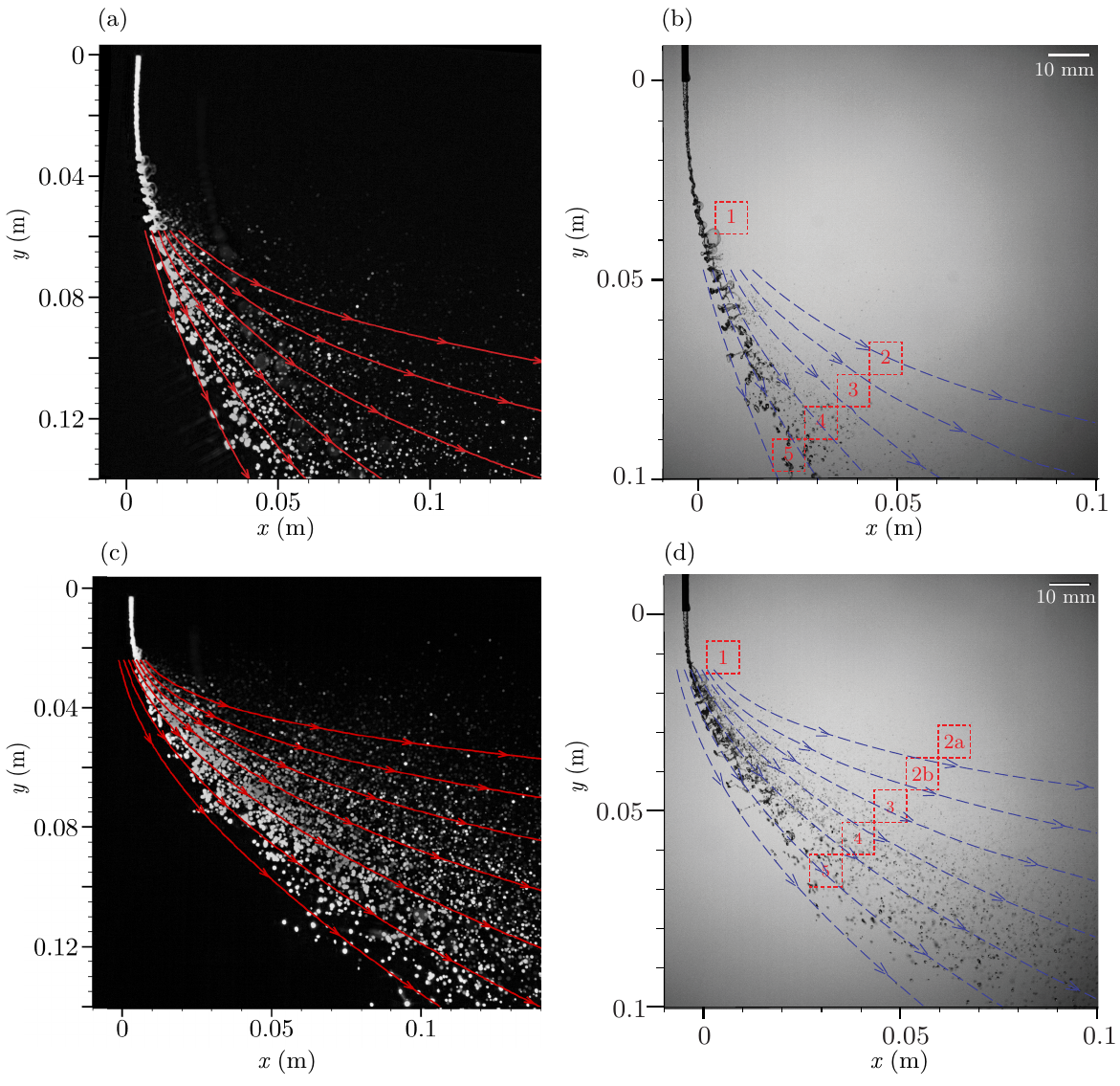}
\caption{Planar Laser-Induced Fluorescence (PLIF) images of a liquid jet undergoing fragmentation for (a,b) $We_j = 1511$ and $We_g = 8$, and (c,d) $We_j = 1511$ and $We_g = 21$. Panels (b) and (d) show the corresponding shadowgraph images, overlaid with the droplet pathlines extracted from the respective PLIF images and the regions selected for the DIH measurements. Panels (a,b) and (c,d) correspond to the same operating conditions as those presented in figures \ref{fig:dsd_We8} and \ref{fig:dsd_We21}, respectively.}
\label{fig:PLIF_shadowgraphy_combined}
\end{figure}

Finally, to determine whether the droplet populations measured by DIH are generated locally or transported from upstream, we perform Planar Laser-Induced Fluorescence (PLIF) experiments for two representative operating conditions. These measurements visualize the liquid-phase evolution and track the trajectories of droplets generated during the different breakup events. A schematic of the PLIF setup used for droplet tracking during the breakup of a liquid jet in a crossflowing airstream is shown in figure \ref{fig:PLIF_schematic}. For the PLIF measurements, Rhodamine 6G dye is dissolved in the 4~L water tank at a concentration of 1\% w/v.  At this low concentration, the addition of Rhodamine 6G does not significantly alter the fluid properties. A dual-cavity pulsed laser (Litron Nano, 200~mJ/pulse, 532~nm wavelength), combined with cylindrical sheet optics ($-25$~mm focal length), is used to form a laser sheet of 0.5~mm thickness, which illuminates the jet in crossflow along the centerline of the test section. Upon excitation at 532~nm, the dye emits fluorescence with a peak wavelength of 550~nm. Accordingly, a 550~nm high-pass filter is fitted to the camera (Phantom VEO 640L, $2560 \times 1600$ pixels). The field of view is approximately 25~cm $\times$ 16~cm, with a spatial resolution of 100.3~$\mu$m/pixel. Dual-frame images are acquired with an inter-frame time of 50~$\mu$s to ensure sufficient droplet displacement between frames, at an acquisition frequency of 10~Hz. Each image pair is then processed using Particle Image Velocimetry (PIV) cross-correlation with a $32 \times 32$ pixel interrogation window and 50\% overlap. A total of 1000 vector fields are averaged to obtain the mean path lines that represent the droplet trajectories. Note that PLIF images are not used to resolve individual droplets. The Rhodamine-dyed droplets serve as tracers, and the PIV analysis provides the representative velocity within each interrogation window. The resulting mean pathlines thus characterize the transport of the droplet population.

The measured droplet trajectories, together with the simultaneous shadowgraph images overlaid with the droplet pathlines extracted from the corresponding PLIF images, are shown in figure~\ref{fig:PLIF_shadowgraphy_combined}(a--d). This approach enables us to identify the origin of the droplets and track their subsequent convective transport along the jet trajectory. The PLIF observations reveal that droplets generated by bag rupture and by bag-rim/node fragmentation follow distinct trajectories immediately after their formation. The fine droplets produced during successive bag ruptures possess a small aerodynamic response time and are therefore rapidly accelerated by the crossflow. Consequently, they are quickly entrained into the gas stream and migrate above the primary jet trajectory. In contrast, the larger droplets generated by bag-rim and node fragmentation possess substantially greater inertia and are less responsive to the aerodynamic forcing. As a result, they remain close to the jet trajectory and continue to penetrate downstream before gradually dispersing. The droplet field, therefore, becomes spatially segregated according to droplet size and breakup origin. Although some downstream transport and mixing inevitably occur, the PLIF observations demonstrate that the characteristic droplet populations measured at the selected DIH locations remain predominantly associated with the corresponding breakup mechanisms discussed above. Thus, while the DIH provides quantitative measurements of droplet size and velocity, the PLIF-based droplet trajectories identify the origin and transport pathways of the droplets in the highly convective jet-in-crossflow.

\section{Concluding remarks} \label{sec:conc}

The fragmentation dynamics of a downward liquid jet subjected to a crossflowing airstream have been investigated experimentally using high-speed shadowgraphy and digital in-line holography (DIH), together with instability-based theoretical modeling. The experimental approach, based on high-speed shadowgraphy and digital in-line holography, enables simultaneous characterization of the jet trajectory, breakup morphology, and volumetric droplet statistics over a broad range of governing parameters. The results demonstrate that the breakup process is governed by the coupled effects of liquid inertia, aerodynamic forcing, and surface tension, as quantified by the liquid and gas Weber numbers and the momentum flux ratio. These parameters determine not only the jet penetration and trajectory but also the transition between distinct breakup regimes, including column, bag, and multibag fragmentation. A semi-empirical model reasonably captures the jet trajectory for different governing parameters. 

A novel outcome of this study is the detailed characterization of droplet size distributions along the jet trajectory using DIH. Our results demonstrate that the breakup process evolves progressively from near-field bag rupture to downstream rim, ligament, and node disintegration, with each fragmentation mechanism associated with distinct characteristic length scales and droplet populations. The measurements reveal a systematic evolution from narrow, fine-droplet distributions in the near field, associated with bag breakup, to broader distributions with larger mean droplet sizes downstream, governed by rim and node disintegration. This progression reflects a cascade of instability mechanisms, primarily Rayleigh–Taylor and Rayleigh–Plateau instabilities, each introducing distinct characteristic length scales into the fragmentation process. We observe that, for the parameters considered, the near-field size distribution is primarily governed by bag breakup. Further downstream, as we progress from the windward region toward the far field, the characteristic droplet sizes are successively dominated by peripheral node breakup, bag-node fragmentation, rim breakup, and, eventually, bag fragmentation. Increasing aerodynamic forcing enhances atomization, producing finer droplets and broader, more polydisperse distributions. This behavior results from reduced instability length scales and an accelerated onset of breakup, leading to a more spatially distributed and dynamically coupled fragmentation process. 

An instability-based theoretical framework is further developed by incorporating the dominant Rayleigh–Taylor and Rayleigh–Plateau instability mechanisms governing bag, rim, and node breakup processes. This framework relates the characteristic instability wavelengths to the resulting droplet size distributions and captures the experimentally observed trends reasonably well across the different fragmentation regimes. The present theoretical framework is developed for moderate Weber-number regimes where bag- and instability-driven breakup mechanisms dominate. At substantially higher gas Weber numbers, where fully shear-driven atomization and turbulent fragmentation become important, additional mechanisms may need to be incorporated into the modeling framework. Furthermore, Planar Laser-Induced Fluorescence (PLIF) measurements are performed to directly visualize droplet trajectories, identify the origin of droplets produced by different breakup modes, and establish the link between the breakup mechanisms and the downstream droplet size distributions. Overall, by comparing theoretical modelling with experimental observations, this work establishes a clear connection between instability-driven breakup mechanisms and evolving droplet size distributions, thereby bridging interfacial dynamics and measurable spray statistics. These findings are relevant to a wide range of applications, including fuel injection, spray cooling, agricultural sprays, and atmospheric dispersion, where accurate characterization of droplet populations is essential. \\

\appendix 
\section{Number-based probability density function ($P_n$)} \label{sec:A1}

\clearpage

\begin{figure}
\centering
\includegraphics[width=0.8\textwidth]{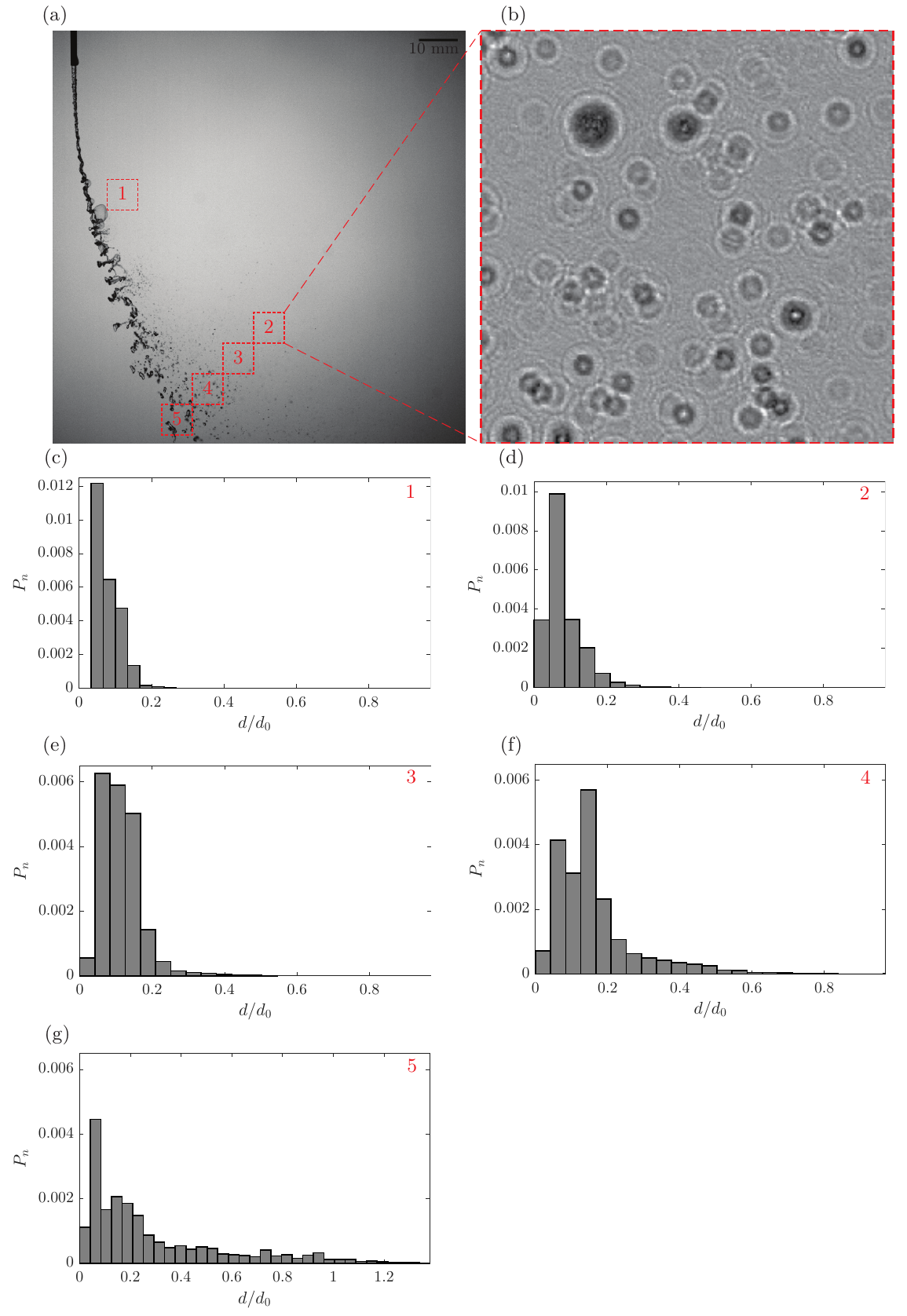}
\caption{Number-based probability density ($P_n$) as a function of the normalized droplet diameter ($d/d_0$) for the five DIH measurement regions 1-5 are shown in panels (c)-(g) for $We_j = 1511$ and $We_g = 8$. This figure corresponds to figure \ref{fig:dsd_We8}. Panels (a) and (b) are identical to those in figure \ref{fig:dsd_We8}(a,b) and are reproduced here for reference.}
\label{fig:npd_We8}
\end{figure}

\begin{figure}
\centering
\includegraphics[width=0.8\textwidth]{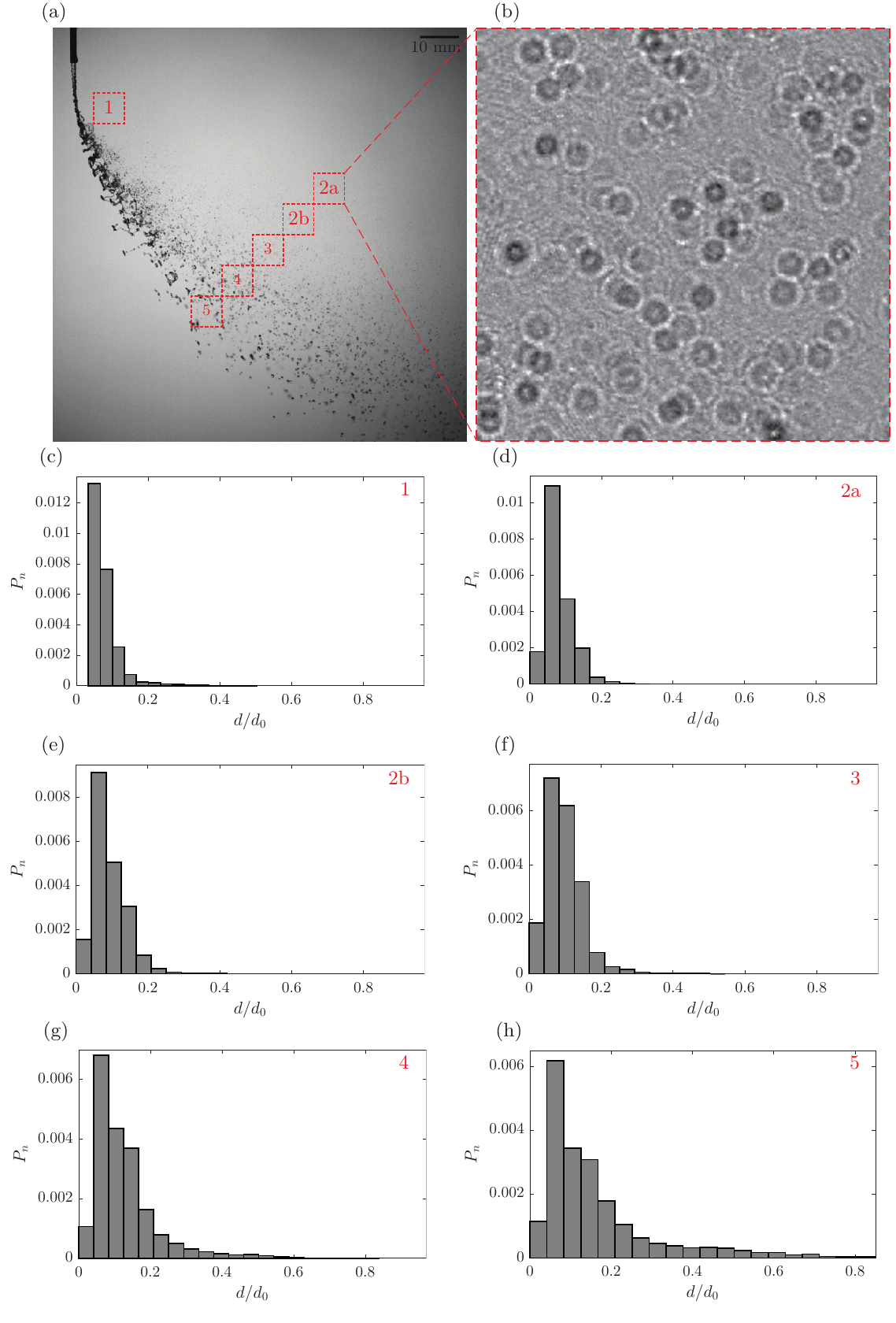}
\caption{Number-based probability density ($P_n$) as a function of the normalized droplet diameter ($d/d_0$) for the five DIH measurement regions 1, 2a, 2b, and 3–5 are shown in panels (c)-(h) for $We_j = 1511$ and $We_g = 21$. This figure corresponds to figure \ref{fig:dsd_We8}. Panels (a) and (b) are identical to those in figure \ref{fig:dsd_We21}(a,b) and are reproduced here for reference.}
\label{fig:npd_We21}
\end{figure}

\clearpage

\noindent{\bf Acknowledgement:} We sincerely thank IIT Hyderabad for providing the necessary support to establish the experimental facility. We sincerely thank the anonymous reviewers for their insightful comments and constructive suggestions, which significantly improved the quality and depth of the manuscript. 
\\

\noindent{\bf Funding:} This work was supported by the Suzuki Next Bharat Fellowship Program (Grant No. NBV/CHE/F011/2025-26/S401). \\

\noindent{\bf Declaration of interests:} The authors report no conflict of interest. \\

\noindent{\bf Data availability statement:} Data will be made available on request. \\

\noindent{\bf Author ORCID:}  \\
G. P. Majji: https://orcid.org/0009-0005-4703-7059 \\
A. J. Tudu: https://orcid.org/0009-0004-6946-5858 \\
L. D. Chandrala: https://orcid.org/0000-0002-2695-9469 \\
K. C. Sahu: https://orcid.org/0000-0002-7357-1141\\

\noindent{\bf Author contributions:} G.P.M. and A.J.T. contributed equally to this study. They established the experimental facility and developed the methodology, conducted the experiments, formulated the theoretical model, and prepared the original draft of the manuscript. L. D. C. and K. C. S. contributed to the supervision, methodology, investigation, funding acquisition, formal analysis, conceptualization, and preparation of the final manuscript. \\


\end{document}